\documentclass[nonacm,acmsmall,screen]{acmart}
\AtBeginDocument{
  }

\usepackage{amsmath,amsfonts}
\usepackage{algorithmic}
\usepackage{algorithm}
\usepackage{array}
\usepackage{subcaption}
\usepackage{textcomp}
\usepackage{stfloats}
\usepackage{url}
\usepackage{verbatim}
\usepackage{todonotes}
\usepackage{pifont}
\usepackage{array}
\usepackage{graphicx}
\usepackage{booktabs}
\usepackage{multicol,multirow}
\usepackage{tabularx}
\usepackage{tikz}
\usetikzlibrary{arrows.meta,positioning}
\usetikzlibrary{shapes.geometric}
\usepackage{pgf,pgfmath,colortbl}
\usepackage{wasysym}
\usepackage{microtype}
\usepackage[most]{tcolorbox}
\usepackage{xparse}
\usepackage[dvipsnames]{xcolor}
\usepackage{pgf}
\usepackage{enumitem}
\usepackage{fontawesome5}

\newcommand{\github}[1]{\faIcon{github}~\url{#1}}

\definecolor{gapgray}{HTML}{F5F6F7}
\definecolor{gapline}{HTML}{6B7280}

\newtcolorbox{gapbox}[1]{
    enhanced,
    colback=gapgray,
    colframe=gapgray,
    boxrule=0pt,
    borderline west={1.2pt}{0pt}{diffpink},
    arc=1.5pt,
    left=6pt,
    right=6pt,
    top=4pt,
    bottom=4pt,
    before skip=4pt,
    after skip=4pt,
    after title={\vspace{-0.4em}},
    fonttitle=\bfseries,
    title={#1},
    coltitle=black
}

\definecolor{promptgray}{RGB}{245,245,245}

\definecolor{diffgreen}{HTML}{117733}
\definecolor{diffgray}{HTML}{BBBBBB}
\definecolor{diffpink}{HTML}{CC6677}
\definecolor{mutedblue}{HTML}{88CCEE}

\newcommand{\diffgreen}[1]{%
    \begingroup
      \setlength{\fboxsep}{0.5pt}%
      \,\colorbox{diffgreen!30}{#1}%
    \endgroup
}
\newcommand{\diffgray}[1]{%
    \begingroup
      \setlength{\fboxsep}{0.5pt}%
      \,\colorbox{diffgray!30}{#1}%
    \endgroup
}
\newcommand{\diffpink}[1]{%
    \begingroup
      \setlength{\fboxsep}{0.5pt}%
      \,\colorbox{diffpink!30}{#1}%
    \endgroup
}

\newtcolorbox{finding}[1]{
    enhanced,
    colback=diffgreen!15,
    colframe=diffgreen!15,
    boxrule=0pt,
    borderline west={1.2pt}{0pt}{diffgreen},
    arc=1.5pt,
    left=6pt,
    right=6pt,
    top=4pt,
    bottom=4pt,
    before skip=4pt,
    after skip=4pt,
    after title={\vspace{-0.4em}},
    fonttitle=\bfseries,
    title={#1},
    coltitle=black
}

\newcommand{\fst}{\cellcolor{diffgreen!30}\bf}  
\newcommand{\worst}{\cellcolor{diffpink!30}}

\DeclareRobustCommand{\sizedots}{%
  \texorpdfstring{%
    \tikz[baseline=-0.55ex]{
      \draw[fill=diffgray!15, draw=black, line width=0.3pt]
        (0,0) circle (0.45ex);
      \draw[fill=diffgray!15, draw=black, line width=0.3pt]
        (1.15ex,0) circle (0.65ex);
      \draw[fill=diffgray!15, draw=black, line width=0.3pt]
        (2.70ex,0) circle (0.80ex);
    }%
  }{model size}%
}

\definecolor{qwen}{HTML}{CC6677}
\definecolor{gemma}{HTML}{332288}
\definecolor{llama}{HTML}{DDCC77}
\definecolor{phi}{HTML}{117733}
\definecolor{deepseek}{HTML}{88CCEE}
\definecolor{mistral}{HTML}{882255}
\definecolor{nextcoder}{HTML}{44AA99}

\newcommand{\modelmarker}[2]{%
  \makebox[2.6mm][c]{%
    \tikz[baseline=-0.55ex]
      \filldraw[
        fill=#1,
        draw=black,
        line width=0.25pt
      ] (0,0) circle (#2);
  }%
}

\begin{document}

\title{Retrieve, Reproduce, Reveal: Dissecting Retrieval-Augmented Software Vulnerability Detection}

\author{Sabrina Kaniewski}
\correspondingauthor
\email{sabrina.kaniewski@hs-esslingen.de}
\orcid{0009-0004-3966-9681}
\affiliation{%
  \institution{Institute for Secure Networked Systems, Esslingen University}
  \city{Esslingen}
  \country{Germany}}

\author{Tim Krämer}
\email{tikrit01@hs-esslingen.de}
\orcid{0009-0004-5231-018X}
\affiliation{%
  \institution{Esslingen University}
  \city{Esslingen}
  \country{Germany}}

\author{Julius Bächle}
\email{julius.baechle@hs-esslingen.de}
\orcid{0009-0009-8997-0129}
\affiliation{%
  \institution{Institute for Intelligent Systems, Esslingen University}
  \city{Esslingen}
  \country{Germany}}
  
\author{Markus Enzweiler}
\email{markus.enzweiler@hs-esslingen.de}
\orcid{0000-0001-9211-9882}
\affiliation{%
  \institution{Institute for Intelligent Systems, Esslingen University}
  \city{Esslingen}
  \country{Germany}}

\author{Michael Menth}
\email{menth@uni-tuebingen.de}
\orcid{0000-0002-3216-1015}
\affiliation{%
  \institution{Chair of Communication Networks, University of Tübingen}
  \city{Tübingen}
  \country{Germany}}

\author{Tobias Heer}
\email{tobias.heer@hs-esslingen.de}
\orcid{0000-0003-3119-252X}
\affiliation{%
  \institution{Institute for Secure Networked Systems, Esslingen University}
  \city{Esslingen}
  \country{Germany}}

\renewcommand{\shortauthors}{Kaniewski et al.}

\makeatletter
\let\@authorsaddresses\@empty
\makeatother

\begin{abstract}
Retrieval-Augmented Generation (RAG) is increasingly used to enhance Large Language Model (LLM)-based software vulnerability detection by grounding predictions in retrieved vulnerability knowledge, such as vulnerability reports.
However, existing RAG-based software vulnerability detection (RAG4SVD) systems are often evaluated using proprietary models, which challenges open science and reproducibility.
Further, studies use different datasets, custom knowledge bases, different backbone models, and diverse metrics, which hinders meaningful cross-system comparison.
In this work, we study six open-source RAG4SVD systems and address these reproducibility and comparability challenges through (i)~reproduction of their experimental settings under an open-weight setting, and (ii)~a unified benchmark using a common dataset, metric suite, and pool of open-weight models.
Further, RAG4SVD systems typically consist of multiple components, yet are often evaluated only as a whole system, i.e., end-to-end. 
Therefore, we perform (iii)~a component-level analysis that decomposes representative RAG4SVD pipelines into input abstraction, knowledge retrieval, and detection.
Our results demonstrate that reproducibility varies substantially across systems.
Under the presented unified benchmark, published RAG4SVD performance does not transfer under a controlled open-weight evaluation and depends strongly on the used model.
The component analysis shows that effective RAG4SVD depends on the alignment between pipeline stages. 
For example, oracle knowledge raises retrieval to near-optimal, yet performance remains low ($0.51$ pairwise accuracy), demonstrating that retrieval effectiveness alone is insufficient for reliable detection.
These findings motivate evaluating RAG4SVD not only end-to-end, but at the level of pipeline components, and provide a basis for more standardized, RAG-aware evaluation practices.
\end{abstract}




\maketitle


\section{Introduction}
\label{sec:introduction}
Large language models (LLMs) are increasingly applied to software vulnerability detection by leveraging their code-understanding and reasoning capabilities to discover insecure code~\cite{kaniewskiLLM4SVD2026}.
However, an LLM's internal knowledge is inherently limited by its training data and respective cutoff date. 
Thus, newly disclosed vulnerabilities, emerging attack patterns, and project-specific security knowledge that were published after the training cutoff date may be unavailable to the model at inference time.
Retrieval-Augmented Generation (RAG)~\cite{lewis2020retrieval} addresses this limitation by dynamically retrieving external vulnerability knowledge, e.g., Common Weakness Enumeration (CWE) descriptions, specific Common Vulnerabilities and Exposures (CVE) reports, or similar vulnerable code examples, from a knowledge base and incorporating it into the prompt~\cite{chenSystematicLiteratureReview}.
Unlike costly retraining or fine-tuning, the knowledge base can be updated independently of the model to add more specific and up-to-date information and, thus, enhance detection performance.

\begin{figure*}[b!]
    \centering
    \includegraphics[width=\columnwidth]{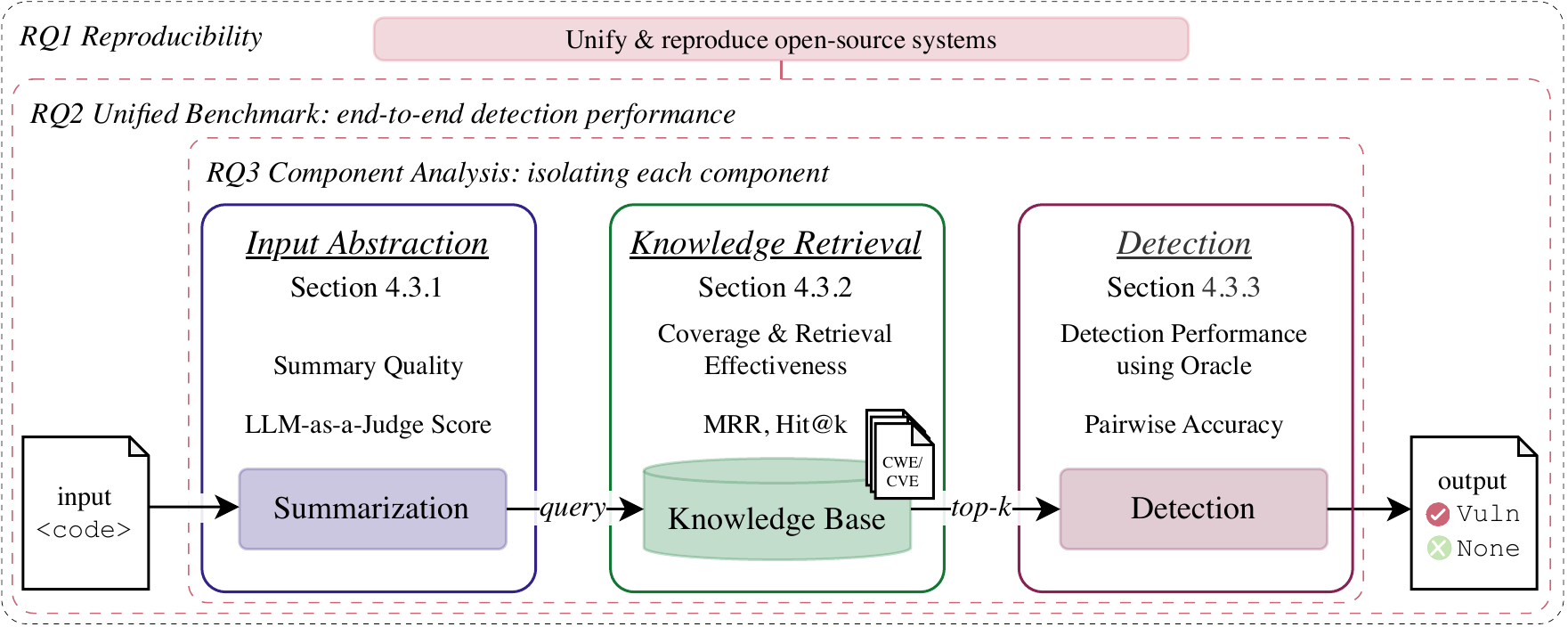}
    \caption{Overview of a typical RAG4SVD pipeline and the evaluation roadmap of this work.}
    \Description{Diagram of a RAG4SVD pipeline from input source code through input abstraction, knowledge retrieval, and vulnerability detection to a vulnerable/non-vulnerable prediction. Nested outlines show how RQ1 evaluates reproducibility, RQ2 evaluates end-to-end detection, and RQ3 separately evaluates abstraction quality, retrieval effectiveness, and detection performance.}
    \label{fig:rag-pipeline}
    \vspace{-0.5em}
\end{figure*}

The adoption of RAG for LLM-based software vulnerability detection (RAG4SVD) has grown rapidly, with proposed RAG systems reporting improvements over non-retrieval baselines~\cite{duVulRAGEnhancingLLMbased2024,kaniewskiLLM4SVD2026,chenSystematicLiteratureReview}.
RAG4SVD systems typically follow a common pipeline, cf. Fig.~\ref{fig:rag-pipeline}:
The code under analysis is first transformed via \textit{input abstraction} into the representation used to query the knowledge base, e.g., the raw code itself or a semantic summary of the code.
This query is passed to a \textit{knowledge retrieval} mechanism, e.g., based on embedding similarity, lexical matching, structured lookup, or a hybrid combination thereof, which retrieves relevant entries.
Finally, the retrieved knowledge is combined with the code under analysis and passed to the LLM for \textit{detection}, producing a vulnerability prediction, e.g., a binary vulnerable/non-vulnerable label.

Yet, it remains unclear how generalizable reported gains are.
First, results are rarely independently verified~\cite{kaniewskiRevisitingVulRAG2026}.
Incomplete public artifacts, undocumented preprocessing and dependencies, non-deterministic inference, and reliance on proprietary models or APIs further complicate reproducing LLM-based systems~\cite{sallouBreakingSilenceThreats2024,AwesomeLLM4SVDTaxonomy,kaniewskiRevisitingVulRAG2026}.
Model providers may discontinue or modify their models without notice, which renders experiments non-reproducible.
Second, systems differ substantially in their datasets, knowledge bases, query representations, retrieval mechanisms, and backbone models~\cite{kaniewskiLLM4SVD2026,chenSystematicLiteratureReview}.
Consequently, results across individual studies do not provide a basis for comparing different systems.
Third, despite the fact that these systems consist of multiple components, they are commonly evaluated only through their final detection accuracy, i.e., end-to-end.
It is often unclear if improvements stem from the quality and coverage of the knowledge base, the abstraction used to formulate a retrieval query, the retrieval mechanism, or the capabilities of the detection model used.
We summarize these observations as three evaluation gaps in the current literature: 

\begin{gapbox}{}
\textbf{\textit{Reproducibility Gap}}: The results of published RAG4SVD systems have rarely been independently reproduced. It remains uncertain whether reported performance can be recovered. 
\end{gapbox}

\begin{gapbox}{}
\textbf{\textit{Comparability Gap}}: Systems are evaluated using different datasets, custom knowledge bases, different models, and diverse metrics, preventing meaningful cross-system comparison.
\end{gapbox}

\begin{gapbox}{}
\textbf{\textit{Attribution Gap}}: RAG systems are predominantly evaluated end-to-end; consequently, it remains unclear which pipeline components drive observed performance improvements.
\end{gapbox}

To address these gaps, we present, to the best of our knowledge, the first study in RAG-based software vulnerability detection to jointly assess reproducibility, cross-system performance, and individual pipeline components.
The \textbf{contributions} of this work are threefold:
(i)~Considering six open-source RAG4SVD systems, i.e., \textit{SVD-Bench}~\cite{zhangBenchmarkingLargeLanguage2025}, \textit{Llama-VD}\footnote{We introduce the abbreviation \textit{Llama-VD} in this work for brevity.}~\cite{ouchebaraLlamabasedSourceCode2025}, \textit{GRACE}~\cite{luGRACEEmpoweringLLMbased2024}, \textit{LLM4Vuln}~\cite{sunLLM4VulnUnifiedEvaluation2024}, \textit{Vul-RAG}~\cite{duVulRAGEnhancingLLMbased2024}, and \textit{VulTriage}~\cite{tangVulTriageTriplePathContext2026}, we first reproduce systems using open-weight models and quantify deviations from their reported results.
(ii)~Second, we analyze how the considered RAG4SVD systems compare when re-implemented under a common dataset, metric suite, and pool of heterogeneous open-weight models, enabling controlled comparison across independently proposed systems.
(iii)~Third, we move beyond end-to-end evaluation and dissect representative multi-stage RAG4SVD pipelines into \textit{input abstraction}, \textit{knowledge retrieval}, and \textit{detection}, proposing a component-aware evaluation of RAG4SVD systems.
Specifically, we evaluate whether generated abstractions preserve information useful for retrieval, whether relevant knowledge is actually retrieved, and whether oracle knowledge improves the detection performance.
Together, these analyses provide a RAG-aware evaluation that examines not \textit{whether} RAG improves vulnerability detection, but \textit{which} components drive performance.
The replication package for this study is publicly available at \github{https://github.com/hs-esslingen-it-security/RAG4SVD}.

The remainder of this work is structured as follows. 
Section~\ref{sec:related_work} reviews related work.
Section~\ref{sec:exp_setup} describes the experimental setup, i.e., the selection of RAG4SVD systems, datasets, metrics, models, and implementation details.
Section~\ref{sec:evaluation} presents the evaluation along the identified gaps and discusses the findings.
We address threats to validity in Sect.~\ref{sec:threats} and conclude this work in
Sect.~\ref{sec:conclusion}.


\section{Related Work}
\label{sec:related_work}
Research on LLM-based software vulnerability detection has expanded rapidly in recent years~\cite{kaniewskiLLM4SVD2026}.
In this section, we review prior work on (i)~benchmarking LLM-based software vulnerability detection, (ii)~benchmarking RAG-based software vulnerability detection approaches, and (iii)~component-level evaluation of RAG pipelines, positioning the contributions of this work.

\subsection{Benchmarking LLM-based Vulnerability Detection}
Several works benchmark how different prompting and tuning strategies affect vulnerability detection performance~\cite{zhangBenchmarkingLargeLanguage2025,steenhoekComprehensiveStudyCapabilities2024,tambergHarnessingLargeLanguage2024,ullahLLMsCannotReliably2024}.
For example, Steenhoek et al.~\cite{steenhoekComprehensiveStudyCapabilities2024} systematically compare prompting strategies, including in-context learning, few-shot, and Chain-of-Thought prompting, and Zhang et al.~\cite{zhangBenchmarkingLargeLanguage2025} compare prompt engineering, instruction tuning, and sequence-classification fine-tuning across programming languages and model sizes.
However, none of these efforts extend to different RAG-based systems, which introduce an additional layer of design choices, such as knowledge base construction or retrieval strategy, on top of existing evaluation challenges.

Recent works further question whether reported performance reflects genuine detection capability of LLMs or is instead an artifact of the datasets and evaluation protocols used~\cite{dingVulnerabilityDetectionCode2024,steenhoekComprehensiveStudyCapabilities2024,ullahLLMsCannotReliably2024}.
Ding et al.~\cite{dingVulnerabilityDetectionCode2024} show that established benchmarks such as BigVul~\cite{BigVul} overestimate model performance due to, e.g., noisy labels and high duplication rates~\cite{guo2023investigation}.
To address these issues, they introduce an improved vulnerability dataset \textit{PrimeVul}, which combines rigorous de-duplication and human-verified labeling with a stricter pairwise evaluation protocol.
Under this more realistic setting, reported F1 scores for state-of-the-art code language models dropped to as low as $0.03$, motivating a shift toward PrimeVul as the de facto benchmark for LLM-based software vulnerability detection, which we also adopt as a common benchmark dataset.
Further, we assess multiple detection models across systems, allowing us to distinguish gains attributable to model capacity from gains attributable to the RAG system architectures.
In doing so, we extend prior LLM benchmarking from model-centric evaluation to system-level evaluation of RAG-based vulnerability detection.

\subsection{Fragmented Evaluation of Retrieval-Augmented Vulnerability Detection}
Existing RAG-based software vulnerability detection systems are predominantly evaluated in isolation, often using custom evaluation protocols, making direct comparisons difficult~\cite{kaniewskiLLM4SVD2026}.

Kaniewski et al.~\cite{kaniewskiRevisitingVulRAG2026} provide an initial reproducibility and replicability study of Vul-RAG~\cite{duVulRAGEnhancingLLMbased2024}.
Motivated by the dependence of many LLM-based approaches on proprietary models, the authors adapt the Vul-RAG pipeline to an open-weight setting and first reproduce its reported open-weight baselines before evaluating a broader set of LLMs.
However, the study scope is intentionally restricted to a single RAG4SVD system and primarily investigates reproducibility and model replicability; it does not compare multiple proposed RAG architectures, transfer them to a new dataset, or evaluate individual pipeline components.

Where multi-system comparisons do exist, they are scoped differently:
For example, VulInstruct~\cite{zhuSpecificationGuidedVulnerabilityDetection2025} and RASM-Vul~\cite{zhaoRetrievalAugmentedSemanticMapping2026a} each compare their detection performance only against Vul-RAG~\cite{duVulRAGEnhancingLLMbased2024}; and VulTriage~\cite{tangVulTriageTriplePathContext2026} ablates pipeline components to quantify their individual contribution.
However, comparisons via final detection performance and ablations do not reveal which RAG design choices generalize, nor whether reported gains reproduce outside their evaluation setting.

This fragmentation is further highlighted by Chen et al.~\cite{chenSystematicLiteratureReview}.
Surveying 30 RAG-based studies across the software vulnerability lifecycle, they highlight that evaluation practices remain fragmented across datasets, baselines, models, and metrics.
We address this gap: to the best of our knowledge, we present the first reproducibility assessment and benchmark for RAG4SVD, applying a common dataset, metric suite, and model pool across systems to make reported gains comparable.

\subsection{RAG Evaluation Methodology and Component Analysis}
RAG evaluation frameworks such as RAGAs~\cite{esRAGAsAutomatedEvaluation2024} and ARES~\cite{saad-falconARESAutomatedEvaluation2024} automate faithfulness and relevance scoring via an LLM-as-a-Judge.
However, these frameworks assume free-text question-answer tasks and do not address the semantics of software vulnerability detection, where retrieval is based on input code and retrieved artifacts consist of code fragments or structured vulnerability knowledge.
For the software vulnerability lifecycle, Chen et al.~\cite{chenSystematicLiteratureReview} further highlight that RAG-specific properties such as retrieval quality, grounding, and faithfulness remain insufficiently evaluated.

Component-wise evaluation of RAG pipelines has been explored for selected software engineering tasks.
Ke et al.~\cite{keNotAllRAGs2026} present a component-wise empirical study that systematically isolates and evaluates four query processing techniques, seven retrieval models, four context refinement methods, and six generators across the code generation, summarization, and repair tasks.
However, their evaluation focuses on extrinsic performance: every component is judged solely by its effect on final task performance, but never assessed individually (e.g., whether retrieved knowledge is relevant).
We address this gap by expanding \textit{extrinsic} evaluation with \textit{intrinsic} evaluation.
Rather than varying components across systems, we evaluate the components of representative RAG4SVD systems in isolation alongside standard detection metrics.

\section{Experimental Setup}
\label{sec:exp_setup}
Our experimental design is built around the three evaluation gaps identified in Sect.~\ref{sec:introduction} and visualized in Fig.~\ref{fig:rag-pipeline}.
To address the \textit{Reproducibility Gap}, we first reproduce considered open-weight systems and assess whether their reported results can be recovered (\textbf{RQ1}).
To address the \textit{Comparability Gap}, we evaluate the systems using a common dataset, metric suite, and open-weight model pool, enabling a controlled comparison of their performance (\textbf{RQ2}).
Finally, to address the \textit{Attribution Gap}, we dissect representative RAG4SVD pipelines into input abstraction, knowledge retrieval, and detection, and evaluate these components individually to determine how they contribute to overall performance (\textbf{RQ3}).
Accordingly, we formulate the following research questions:
\begin{enumerate}[label={\small\textbf{RQ\arabic*}}, leftmargin=3.5em, labelsep=0.75em]
\setlength{\itemsep}{0.1em}
    \item To what extent are published open-source RAG4SVD systems reproducible?
    \item How do open-source RAG4SVD systems compare under a unified benchmark?
    \item Which component-level factors drive detection in representative RAG4SVD pipelines?
\end{enumerate}


\subsection{RAG4SVD System Selection}
\label{sec:rag_selection}

\begin{figure*}[b!]
    \centering
    \begin{tikzpicture}[
        >=Latex,
        font=\fontsize{7pt}{7pt}\selectfont,
        box/.style={
            draw,
            rounded corners=3pt,
            align=center,
            minimum height=.9cm,
            inner xsep=8pt,
            inner ysep=4pt
        },
        flowlabel/.style={
            font=\fontsize{7pt}{7pt}\selectfont,
            inner sep=1pt,
            align=center,
            yshift=2pt
        }
    ]

    \node[box] (scope) {\textbf{31}\\RAG4SVD studies~\cite{AwesomeLLM4SVDTaxonomy}};
    \node[box, right=2.2cm of scope] (artifacts) {\textbf{12}\\studies with public artifacts};
    \node[box, right=2.2cm of artifacts] (repro) {\textbf{6}\\executable systems};

    \draw[->, thick] (scope.east) -- node[flowlabel, above] {artifact availability} (artifacts.west);
    \draw[->, thick] (artifacts.east) -- node[flowlabel, above] {completeness \&\\documentation} (repro.west);
    \vspace{-0.5em}
    \end{tikzpicture}
    \caption{RAG4SVD system screening and selection.}
    \Description{Flow diagram of the study-selection process. From 31 identified RAG4SVD studies, 12 provide publicly accessible code or artifacts. Of these, 6 provide a sufficiently complete and documented setup, including the pipeline, data, and experimental configuration, to support reproduction.}
    \label{fig:selection_flow}
\end{figure*}
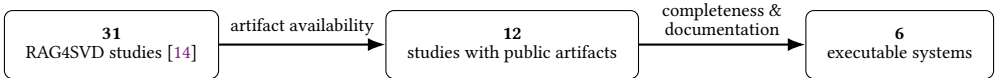

We consider RAG as the dynamic selection of external context at inference-time, conditioned on the code under analysis.
Within RAG4SVD, we consider \textit{example-retrieval RAG}, where semantically or lexically similar labeled code examples are selected for few-shot prompting, and \textit{knowledge-retrieval RAG}, where the system queries a dedicated vulnerability knowledge base. 

A continuously updated survey artifact by Kaniewski et al.~\cite{AwesomeLLM4SVDTaxonomy} tracks LLM-based software vulnerability detection literature. 
As of August 2026, it lists $31$ studies implementing RAG within the RAG4SVD scope.
For this study, we require (i) publicly accessible implementation code or equivalent executable artifacts and (ii) a sufficiently complete and documented setup to reconstruct the RAG pipeline and its evaluation, including the required data, configuration, and reported experimental settings.
Applying these criteria leaves $12$ ($39\%$) studies with public artifacts, of which only $6$ provide a complete and executable setup for reproduction.
Fig.~\ref{fig:selection_flow} summarizes the selection process.
We provide study-specific inclusion and exclusion rationale in the replication package.
Tab.~\ref{tab:rag4svd_systems} summarizes the selected systems along the RAG4SVD pipeline stages.

\newcommand{\included}{\diffgreen{\textcolor{diffgreen}{\ding{51}}}}
\begin{table*}[b]
    \centering
    \fontsize{7pt}{7pt}\selectfont
    \setlength{\tabcolsep}{3pt}
    \caption{Overview of representative RAG4SVD systems considered in this study, characterized by their retrieval query, retrieval mechanism, retrieved knowledge, and coverage in the research questions RQ1--RQ3.}
    \label{tab:rag4svd_systems}
    \begin{tabular}{@{}llll>{\centering}p{0.46cm}>{\centering}p{0.46cm}>{\centering\arraybackslash}p{0.46cm}@{}}
        \toprule
        \textbf{System} &
        \textbf{Query} &
        \textbf{Retrieval} &
        \textbf{Retrieved Knowledge} &
        \textbf{RQ1} &
        \textbf{RQ2} &
        \textbf{RQ3} \\
        \midrule

        \textbf{SVD-Bench}~\cite{zhangBenchmarkingLargeLanguage2025}
        & Code
        & Dense (\texttt{simcse-bert-base})
        & Code examples
        & \included & \included & \\

        \textbf{Llama-VD}~\cite{ouchebaraLlamabasedSourceCode2025}
        & Code
        & Dense (\texttt{codebert-base})
        & Code examples
        & \included & \included & \\

        \textbf{GRACE}~\cite{luGRACEEmpoweringLLMbased2024}
        & Code
        & Dense (\texttt{CodeT5}) + re-ranking
        & Code examples
        & & \included & \\

        \textbf{LLM4Vuln}~\cite{sunLLM4VulnUnifiedEvaluation2024}
        & Semantic abstr.
        & Dense (\texttt{text-embedding-ada-002})
        & Summarized CWE reports
        & \included & & \included \\

        \textbf{Vul-RAG}~\cite{duVulRAGEnhancingLLMbased2024}
        & Code + semantic abstr.
        & Sparse (BM25) + re-ranking
        & Functionality + root causes + patches
        & \included & \included & \included \\

        \textbf{VulTriage}~\cite{tangVulTriageTriplePathContext2026}
        & Vulnerability abstr.
        & Hybrid (dense + sparse; \texttt{bge-m3})
        & CWE descriptions + code examples
        & & \included & \\
        
        \bottomrule
    \end{tabular}
\end{table*}

\textit{SVD-Bench~\cite{zhangBenchmarkingLargeLanguage2025}} studies RAG as one of several prompting techniques. 
It uses dense retrieval, which embeds the function under analysis and labeled training functions with the \texttt{simcse-bert-base} model and ranks them by cosine similarity to retrieve the top-$n$ most similar examples.

\textit{Llama-VD~\cite{ouchebaraLlamabasedSourceCode2025}} investigates different prompting and fine-tuning strategies for adapting Llama-3.1-8B to vulnerability detection. 
It applies dense retrieval using \texttt{codebert-base}, and the $6$ nearest training examples are retrieved using Euclidean distance. 
The study additionally explores using RAG-selected examples in test-time tuning, thereby combining retrieval with parameter adaptation.

\textit{GRACE~\cite{luGRACEEmpoweringLLMbased2024}} uses dense retrieval, embedding the function under analysis and training functions with \texttt{CodeT5} to retrieve semantically similar candidates.
These candidates are subsequently re-ranked using lexical token overlap and syntactic similarity derived from Abstract Syntax Tree (AST) representations to select a single example.
GRACE further extracts structural program information from the code and supplies it alongside the code and retrieved example in the prompt.

\textit{LLM4Vuln~\cite{sunLLM4VulnUnifiedEvaluation2024}} presents a modular evaluation framework that separates an LLM's intrinsic vulnerability reasoning from external enhancements, including knowledge retrieval, program-context supplementation, and prompting.
For knowledge retrieval, it uses dense retrieval with \texttt{text-embedding-ada-002}, supporting either the raw source code or an LLM-generated functionality summary (semantic abstraction) as the query representation.
The top-$3$ most similar entries are retrieved, and their associated raw or summarized vulnerability knowledge, including vulnerability descriptions and root causes, is supplied to the LLM for detection.

\textit{Vul-RAG~\cite{duVulRAGEnhancingLLMbased2024}} retrieves abstract, vulnerability-specific knowledge distilled from vulnerable-patched code pairs.
Its knowledge base represents each vulnerability through functional semantics, vulnerability causes, and corresponding patches.
At inference, an LLM generates an abstract purpose and detailed behavioral description of the code under analysis; Vul-RAG uses these together with the raw code as three retrieval queries.
It applies sparse BM25 retrieval, which ranks entries by lexical term matching, and combines the resulting candidate lists using reciprocal rank fusion.
The retrieved items are then examined sequentially by the LLM to determine whether the code exhibits a retrieved vulnerability cause without the corresponding patch behavior.

\textit{VulTriage~\cite{tangVulTriageTriplePathContext2026}} augments vulnerability detection through three complementary context paths targeting program structure, program semantics, and vulnerability knowledge.
For knowledge retrieval, it generates up to two coarse natural-language vulnerability predictions of the code under analysis in a first step and uses these predictions as queries to a CWE-derived knowledge base.
It performs hybrid retrieval by combining dense semantic similarity using \texttt{bge-m3} embeddings with sparse lexical matching.
The retrieved CWE descriptions and code examples are then combined with the code under analysis, structural context, and semantic summary for the final prediction.


\subsection{Datasets}
\label{sec:datasets}
In the evaluation, we distinguish between the original datasets used by each vulnerability detection system and a common dataset used for cross-system comparison. 
For RQ1, we retain each system's original dataset and data split (including the system-specific knowledge base) to reproduce its reported performance.
For the unified benchmark~(RQ2), we use PrimeVul~\cite{dingVulnerabilityDetectionCode2024} as the common dataset.
PrimeVul was introduced to address limitations of earlier vulnerability datasets, particularly noisy labels, duplicate samples, and data leakage across training and evaluation sets~\cite{dingVulnerabilityDetectionCode2024}. 
The full dataset~(v0.1) contains 6,968 vulnerable and 228,800 benign functions spanning 94 CWEs and employs a chronological train, validation, and test split.
For the benchmark, we specifically use the paired subset of PrimeVul, i.e., \textbf{PrimeVul Paired}: each vulnerable function is matched with its corresponding patch, resulting in 5480 pairs across 88 CWEs, making it particularly suitable for evaluating whether a model can correctly distinguish vulnerable code from its patched counterpart. We use the train split as the basis for the knowledge base and the test split for testing.
For the component analysis in RQ3, specifically the knowledge retrieval, we additionally construct test set probes, i.e., randomly sample from the test sets (see Appendix~\ref{app:datasets}).

\subsection{Metrics}
\label{sec:metrics}
\textit{Detection Performance.}
We assess detection performance using the standard classification metrics: Accuracy, Precision, Recall, and F1-score. 
While these metrics quantify performance at the level of individual functions, they do not capture whether a model can distinguish a vulnerable function from its patched non-vulnerable counterpart.
Following the pairwise evaluation protocol introduced for PrimeVul~\cite{dingVulnerabilityDetectionCode2024}, we report \textbf{Pairwise Accuracy}~\cite{duVulRAGEnhancingLLMbased2024}. 
A vulnerability-patch pair is counted as correctly classified only if the vulnerable instance is predicted as vulnerable \textit{and} its corresponding patched instance is predicted as non-vulnerable:
$
\mathrm{Pair.\,Acc.}
=
\frac{1}{N}
\sum_{i=1}^{N}
\mathbf{1}
(
\hat{y}^{\mathrm{vuln}}_i=1
\land
\hat{y}^{\mathrm{patch}}_i=0
)
$,
where $N$ denotes the number of pairs.
Since paired instances differ primarily in the security-relevant changes introduced by the patch, this metric provides a stricter test of whether a model correctly distinguishes vulnerable from patched code.

\textit{Retrieval Effectiveness.}
For evaluating retrieval effectiveness, we adopt \textbf{Hit@$k$} and Mean Reciprocal Rank~(\textbf{MRR})~\cite{gan2025retrievalaugmentedgenerationevaluation}, treating a retrieved item as a successful match if it shares the queried code's ground truth CWE, considering both exact matches and hierarchical CWE parent/child relaxations (cf. Tamberg and Bahsi~\cite{tambergHarnessingLargeLanguage2024}).
Hit@$k$ measures the proportion of queries for which at least one successful match occurs within the top-$k$ retrieved items:
$
\mathrm{Hit@}k = \frac{1}{N}\sum_{i=1}^{N} \mathbf{1}(r_i \leq k)
$,
where $N$ is the number of queries and $r_i$ denotes the rank of the first successful match for query $i$.
MRR captures how high the first successful match is ranked, averaged across all queries:
$
\mathrm{MRR} = \frac{1}{N}\sum_{i=1}^{N} \frac{1}{r_i}
$.

\subsection{Model Selection}
\label{sec:models}
Several of the considered RAG4SVD systems originally rely on proprietary models, e.g., GPT-4~\cite{gpt4}, or use open-weight models of scales typically requiring API services, e.g., DeepSeek-V3~\cite{deepseekv3technicalreport} with $685$B parameters.
For the unified evaluation, we select a common pool of $22$ open-weight models from $7$ model families: Qwen~\cite{qwen25technicalreport,qwen25codertechnicalreport,qwen3technicalreport,qwen35blog,qwen36blog,qwq32b}, Llama~\cite{llama3}, Phi~\cite{phi4technicalreport,phi4minitechnicalreport}, Gemma~\cite{gemma3technicalreport}, NextCoder~\cite{nextcoder}, DeepSeek~\cite{deepseekr1}, and Mistral~\cite{mistral7b}.
The pool spans parameter scales from approximately $3$B to $32$B, deliberately focusing on scales that remain amenable to practical local and on-premises deployment. 
The selection further includes both general-purpose and code-specialized models. 
Where available, we include multiple parameter scales from the same family, allowing us to additionally examine the effect of model scale.
Unless otherwise stated, we use the full model pool consistently across experiments. 
The complete model lineup and checkpoints are reported in Appendix~\ref{app:model-lineup}.

\subsection{Implementation Details}
\label{sec:implementation}
To enable a consistent evaluation across systems, we integrate all considered RAG systems into a unified local inference pipeline using Hugging Face\footnote{\url{https://huggingface.co}, accessed 11-09-2026.} and the Transformers library. 
Systems that originally relied on API-based models were adapted by replacing their API calls with a common interface for locally hosted open-weight models.
We preserve all other system logic, prompts, retrieval configurations, and system-specific hyperparameters.

Minor implementation changes were necessary to accommodate differences between proprietary and open-weight models. 
In particular, some open-weight models do not always adhere as reliably to requested output formats, e.g., requiring \texttt{<result>YES/NO<\textbackslash result>} tags~\cite{duVulRAGEnhancingLLMbased2024}, which requires more robust parsing of generated responses. 
We therefore hardened system-specific output parsers where necessary to tolerate formatting variations without changing the semantic interpretation of model outputs. 
Detailed system-specific changes are documented in the replication package.

All retrieval and detection experiments with locally hosted open-weight models were executed on a high-performance computing cluster equipped with NVIDIA L40S 48\,GB and NVIDIA H100 80\,GB GPUs.
The only exception to local inference is the LLM-as-a-Judge analysis in RQ3.
Following the judge quality assessment by Li et al.~\cite{liSFTRLDemystifying2026}, we use GPT-oss-120B~\cite{gptoss120b} as the judge model and access it via the Blablador API\footnote{The Blablador API, provided by Jülich Supercomputing Centre, is accessible to researchers: \url{https://sdlaml.pages.jsc.fz-juelich.de/ai/guides/blablador_api_access/}, accessed 11-09-2026.} due to its computational requirements.
The judge is used only for evaluating generated abstractions and does not participate in retrieval or vulnerability detection.


\section{Evaluation}
\label{sec:evaluation}
In this section, we report the empirical evaluation of the considered RAG4SVD systems, structured around the derived research questions RQ1-RQ3.
We first determine whether reported results can be reproduced (RQ1), then compare reproducible systems using a unified benchmark (RQ2), and finally isolate individual RAG components (RQ3).

\subsection{RQ1 Reproducibility}
\label{sec:rq1}
Reproducibility is a prerequisite for meaningful benchmarking.
If systems cannot be faithfully reproduced, differences observed in subsequent comparisons may reflect implementation artifacts rather than underlying methods.
For RQ1, we focus on systems that report results for at least one open-weight model.
This criterion includes SVD-Bench, Llama-VD, LLM4Vuln, and Vul-RAG, while excluding GRACE and VulTriage, whose evaluations rely exclusively on proprietary models.

\begin{figure*}[b!]
    \centering
    \includegraphics[width=\columnwidth, trim=0 7pt 0 0, clip]{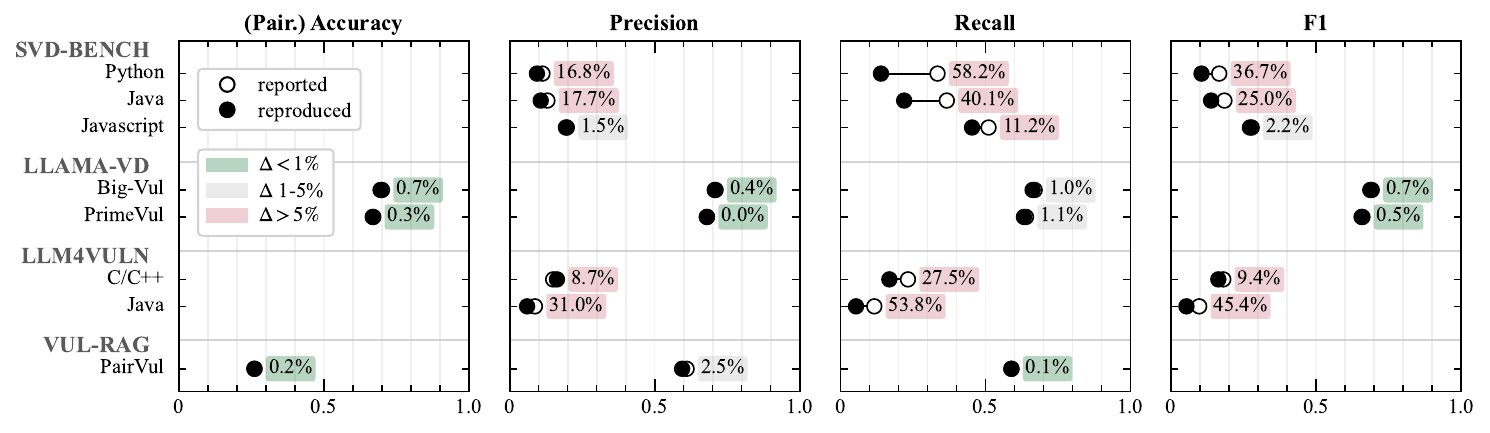}
    \caption{Reproducibility of considered RAG4SVD systems. Absolute relative deviation between $\circ$~reported and $\bullet$~reproduced results for each metric reported, averaged across all evaluated open-weight models. Lower values indicate higher reproducibility; deviations below $1\%$ are considered \diffgreen{reproducible}, $1$-$5\%$ \diffgray{weakly reproducible}, and above $5\%$ \diffpink{not reproducible}. See Appendix~\ref{app:reproducibility} for the absolute differences.}
    \Description{Dumbbell plots compare originally reported and reproduced performance for SVD-Bench, Llama-VD, LLM4Vuln, and Vul-RAG across accuracy, precision, recall, and F1. Relative deviations are annotated and categorized as reproducible, weakly reproducible, or not reproducible.}
    \label{fig:reproduced_results}
\end{figure*}

For the considered systems, we assess whether the reported results can be \textit{reproduced in a local open-weight inference setting}.
Where a study evaluates multiple datasets or models, we reproduce these settings separately.
We repeat each experiment three times to account for inference variability; results are averaged across runs.
Following Nong et al.~\cite{nongopenscience}, we quantify reproducibility using the absolute relative deviation between reproduced and reported results.
Deviations below $1\%$, between $1\%$ and $5\%$, and above $5\%$ are classified as reproducible, weakly reproducible, and not reproducible, respectively.
Fig.~\ref{fig:reproduced_results} summarizes the results across all considered systems.

\textit{SVD-Bench.}
We reproduce the results of SVD-Bench across its Python, Java, and JavaScript evaluation settings. 
Reproducibility does not hold consistently across this language coverage: deviations exceed $5\%$ for Python and Java, while selected JavaScript metrics fall within the weakly reproducible range.
The largest deviation occurs for recall on the Python dataset ($58.2\%$), indicating that the reproduced implementation does not recover the reported performance. 

\textit{Llama-VD.}
We reproduce the results of Llama-VD for Llama-3.1-8B on both originally evaluated datasets, Big-Vul and PrimeVul. 
Reproducibility is consistent across these datasets, with nearly all reported metrics deviating by less than $1\%$; only recall marginally exceeds this threshold ($1.0$-$1.1\%$).

\textit{LLM4Vuln.}
LLM4Vuln reports results for different detection models, including three open-weight models, but still relies on GPT-4.1 for summary generation and output evaluation and, thus, cannot be reproduced with its original configuration in a fully open-weight reproduction setting.
We therefore substitute these components with open-weight alternatives.
We chose the substitutes (summary: Qwen3-4B, judge: Gemma-3-27B) through an ablation of candidate models based on their error rate and ability to produce correctly parseable outputs (cf. replication package). 
Despite using the closest-performing substitutes, the reproduced results deviate by more than $5\%$ from the reported results, rendering LLM4Vuln not reproducible under a fully open-weight setting.
Unlike the other considered systems, this deviation is not attributable to, e.g., parsing or inference variance alone, since LLM4Vuln depends on unavailable proprietary components.
The reported results here reflect a model substitution rather than a reproduction of the original configuration, making a deviation from the reported results expected.

\textit{Vul-RAG.}
We reproduce the results of Vul-RAG for Qwen2.5-Coder-32B on the PairVul dataset.
The reproduced results closely match the reported performance, with deviations below $1\%$ for accuracy and recall; precision exhibits a small deviation of $2.5\%$, i.e., is weakly reproducible.

\begin{finding}{}
\textbf{Finding \#1}: 
Reproducibility varies substantially across the considered RAG4SVD systems, with potential contributing factors including parser hardening and inference non-determinism.
For systems that require substituting proprietary models, reproducibility in an open-weight setting is unattainable.
\end{finding}


\subsection{RQ2 Unified Benchmark}
\label{sec:rq2}

\begin{table*}[b!]
    \centering
    \fontsize{7pt}{7pt}\selectfont
    \setlength{\tabcolsep}{3pt} 
    \caption{Performance comparison across RAG4SVD systems on the PrimeVul Paired dataset. 
    We report a cell only if at least 95\% of pairs yielded a correctly parsed prediction.
    \colorbox{diffgreen!30}{\textbf{Best}} and \colorbox{diffpink!30}{worst} pairwise accuracy per system highlighted. 
    For the best configuration per system, we repeat inference $3\times$ and report mean $\pm$ std.
    The system average is computed over the intersection of models (marked with *) that yielded a prediction for \textit{all} systems, and over the top-$5$ performing models per system.
    }
    \label{tab:benchmark}
    \begin{tabular}{@{}lcccccccccc@{}}
        \toprule
        & \multicolumn{2}{c}{\textbf{SVD-Bench}}
        & \multicolumn{2}{c}{\textbf{GRACE}}
        & \multicolumn{2}{c}{\textbf{Llama-VD}}
        & \multicolumn{2}{c}{\textbf{Vul-RAG}}
        & \multicolumn{2}{c}{\textbf{VulTriage}} \\
        \cmidrule(lr){2-3}
        \cmidrule(lr){4-5}
        \cmidrule(lr){6-7}
        \cmidrule(lr){8-9}
        \cmidrule(lr){10-11}
        \textbf{LLM}
        & Pair. Acc. & F1
        & Pair. Acc. & F1
        & Pair. Acc. & F1
        & Pair. Acc. & F1
        & Pair. Acc. & F1 \\
        \midrule



        
        Qwen2.5-Coder-3B
        & 0.07 & 0.33
        & 0.04 & 0.62
        & 0.07 & 0.45
        & - & -
        & - & - \\

        Qwen2.5-Coder-14B*
        & 0.09 & 0.48
        & 0.02 & 0.39
        & 0.07 & 0.46
        & \fst 0.22 $\pm$ 0.01 & 0.55
        & 0.07 & 0.14 \\

        Qwen2.5-Coder-32B*
        & 0.09 & 0.43
        & 0.04 & 0.39
        & 0.11 & 0.48
        & 0.16 & 0.53
        & \fst 0.21 $\pm$ 0.00 & 0.37 \\

        Qwen2.5-3B
        & 0.13 & 0.49
        & 0.10 & 0.56
        & 0.10 & 0.46
        & 0.12 & 0.51
        & - & - \\

        Qwen2.5-14B
        & \fst 0.12 $\pm$ 0.01 & 0.53
        & 0.04 & 0.46
        & 0.09 & 0.49
        & 0.18 & 0.55
        & - & - \\

        Qwen2.5-32B
        & 0.12 & 0.56
        & 0.03 & 0.44
        & 0.13 & 0.48
        & 0.16 & 0.56
        & - & - \\

        Qwen3-4B
        & 0.10 & 0.47
        & 0.10 & 0.26
        & 0.09 & 0.47
        & 0.17 & 0.54
        & - & - \\

        Qwen3.5-9B
        & - & -
        & 0.05 & 0.17
        & \fst 0.15 $\pm$ 0.00 & 0.50
        & - & -
        & - & - \\

        Qwen3.6-27B
        & - & -
        & 0.01 & 0.01
        & 0.06 & 0.42
        & 0.14 & 0.56
        & - & - \\

        QwQ-32B
        & 0.12 & 0.58
        & \fst 0.17 $\pm$ 0.00 & 0.62
        & \worst 0.06 & 0.43
        & 0.21 & 0.57
        & - & - \\

        Llama-3.2-3B
        & 0.09 & 0.37
        & 0.01 & 0.67
        & 0.12 & 0.48
        & - & -
        & - & - \\

        Llama-3.1-8B
        & 0.12 & 0.46
        & \worst 0.00 & 0.67
        & 0.12 & 0.49
        & \worst 0.12 & 0.53
        & - & - \\

        Phi-4-mini
        & 0.11 & 0.51
        & 0.09 & 0.55
        & 0.09 & 0.46
        & - & -
        & - & - \\

        Phi-4
        & 0.10 & 0.36
        & 0.04 & 0.42
        & 0.11 & 0.49
        & 0.14 & 0.56
        & - & - \\

        Gemma-3-4B
        & 0.10 & 0.47
        & 0.07 & 0.64
        & 0.15 & 0.51
        & 0.19 & 0.50
        & - & - \\

        Gemma-3-12B*
        & 0.08 & 0.36
        & 0.04 & 0.49
        & 0.09 & 0.48
        & 0.20 & 0.54
        & \worst{0.06} & 0.63 \\

        Gemma-3-27B*
        & \worst 0.07 & 0.38
        & 0.03 & 0.58
        & 0.08 & 0.48
        & 0.20 & 0.55
        & 0.21 & 0.70 \\

        NextCoder-14B*
        & 0.08 & 0.43
        & 0.02 & 0.40
        & 0.07 & 0.46
        & 0.22 & 0.56
        & 0.07 & 0.13 \\

        NextCoder-32B
        & 0.12 & 0.51
        & 0.06 & 0.40
        & 0.07 & 0.48
        & 0.19 & 0.55
        & - & - \\

        DeepSeek-R1-8B
        & 0.12 & 0.48
        & 0.14 & 0.58
        & 0.09 & 0.46
        & 0.21 & 0.51
        & - & - \\

        DeepSeek-R1-32B
        & 0.10 & 0.50
        & 0.16 & 0.51
        & 0.10 & 0.49
        & 0.19 & 0.53
        & - & - \\

        Mistral-v0.3-7B
        & 0.09 & 0.41
        & 0.06 & 0.57
        & 0.14 & 0.49
        & - & -
        & - & - \\

        \midrule
        \textbf{Average ($n=5$, $*$intersection)}
        & 0.08 & 0.41
        & 0.03 & 0.45
        & 0.08 & 0.47
        & 0.20 & 0.55
        & 0.12 & 0.40 \\

        \textbf{Average ($n=5$, best)}
        & 0.12 & 0.53
        & 0.13 & 0.51
        & 0.14 & 0.49
        & 0.21 & 0.55
        & 0.12 & 0.40 \\
        \bottomrule
    \end{tabular}
\end{table*}

We evaluate the considered systems under a common dataset, metric suite, and model pool to enable a controlled cross-system comparison.
Notably, we are not estimating each system's best possible performance; we assess how the systems transfer under a controlled open-weight environment.
Accordingly, we exclude LLM4Vuln from the unified benchmark. 
LLM4Vuln's modular pipeline relies on GPT-4.1 for summarization and judge-based evaluation such that evaluating it in an open-weight inference setting would require model substitutions, which introduces confounds beyond the choice of detection backbone.
To further separate detection performance from parsing failures and prompt-specific instruction-following issues, we report a configuration only if at least 95\% of pairs yield a correctly parsed prediction.

\textit{System Ranking.} 
Table~\ref{tab:benchmark} summarizes the performance across systems and backbone models.
Averaged over the intersection of models that yielded a prediction for \textit{all} systems, Vul-RAG achieves the highest mean pairwise accuracy~($0.20$), followed by VulTriage~($0.12$), SVD-Bench and Llama-VD~($0.08$), and GRACE~($0.03$). 
Notably, SVD-Bench and Llama-VD, which both implement example-retrieval RAG using dense retrieval (with different embedding models), converge to the same mean pairwise accuracy.
Considering the top-$5$ best performing models per system, the ordering slightly shifts: Vul-RAG reaches $0.21$ pairwise accuracy, followed by Llama-VD~($0.14$), GRACE~($0.13$), SVD-Bench~($0.12$), and finally VulTriage~($0.12$).
Considering the {\setlength{\fboxsep}{1pt}%
\colorbox{diffgreen!30}{\textit{strongest individual configurations}}}, we repeat each configuration three times to assess inference variability and report the mean pairwise accuracy $\pm$ standard deviation.
Vul-RAG reaches $0.22 \pm 0.01$, compared with $0.21 \pm 0.00$ for VulTriage, $0.17 \pm 0.00$ for GRACE, $0.15 \pm 0.00$ for Llama-VD, and $0.12 \pm 0.01$ for SVD-Bench.
The low standard deviations indicate limited run-to-run inference variability; we therefore report single-run results for the remaining benchmark configurations.

Pairwise accuracy also reveals failure modes that are obscured by F1 score alone.
For example, Llama-3.2-3B and Llama-3.1-8B under GRACE achieve F1 scores of $0.67$, while their pairwise accuracy is only $0.01$ and $0.00$, respectively.
For the pairwise evaluation setting, this combination is attributable to almost always predicting \texttt{vulnerable}. 
Despite a seemingly strong F1 score, the system fails to distinguish vulnerable functions from their patched counterparts.

\textit{Backbone Sensitivity.}
Performance varies markedly with the underlying detection model, but we observe no consistent advantage through model scale or code specialization where model families match.
Within the Qwen2.5 family, for example, performance does not increase monotonically with parameter scale across systems, and the corresponding coder variants do not systematically outperform their general-purpose counterparts.
Instead, the strongest backbone is pipeline-specific.

\textit{Parsing Robustness.}
The benchmark additionally exposes substantial differences in output-format robustness across systems.
GRACE and Llama-VD yield sufficiently parseable predictions across the evaluated model pool. 
In contrast, VulTriage, which reports SOTA pairwise accuracy of $0.34$ using GPT-4o on PrimeVul Paired~\cite{tangVulTriageTriplePathContext2026}, meets the 95\% parsing threshold for only $5$ tested open-weight models, indicating limited transferability of its prompting and output-format beyond the original proprietary-model setting.
Instruction following and detection performance should therefore be treated as distinct system properties: a pipeline may perform well for a compatible backbone while remaining difficult to evaluate reliably with others due to prompt or output-format sensitivity.


\begin{finding}{}
\textbf{Finding \#2}: 
Using a common benchmark, RAG4SVD systems differ substantially in detection performance, backbone sensitivity, and parsing robustness.
Vul-RAG achieves the highest mean pairwise accuracy, while reported SOTA rankings under the systems' original evaluation settings do not transfer to a common open-weight setting.
These results show that system-level gains strongly depend on the backbone model and evaluation setting used.
\end{finding}


\subsection{RQ3 Component Analysis}
RAG4SVD pipelines combine multiple components, e.g., input abstraction, knowledge retrieval, and retrieval-augmented detection, as shown in Fig.~\ref{fig:rag-pipeline}.
However, prior work typically reports only end-to-end detection performance, leaving open which components drive the observed improvements.
We complement the \textit{extrinsic} evaluation in RQ2 with an \textit{intrinsic} analysis that isolates and evaluates individual pipeline components.
For this analysis, we focus on Vul-RAG~\cite{duVulRAGEnhancingLLMbased2024} and LLM4Vuln~\cite{sunLLM4VulnUnifiedEvaluation2024}, as both systems expose separable abstraction, retrieval, and detection stages and use retrieved knowledge as primary external augmentation, which facilitates component-level attribution.


\subsubsection{Input Abstraction}
\label{sec:rq3-summary}
We first evaluate the quality of input abstractions, i.e., whether using summaries of code functionality preserves information necessary for retrieval.
Following prior work on LLM-based code summarization evaluation~\cite{wuCanLargeLanguage2025,wangCanLLMsReplace2025,sunSourceCodeSummarization2025,songFineSurEFinegrainedSummarization2024,crupiEffectivenessLLMasaJudgeCode2025}, we use an LLM-as-a-Judge and assess each summary along five dimensions.
Four of these dimensions capture established summary-quality properties: \textit{Faithfulness} (support by the input code), \textit{Functional Coverage} (coverage of the code's primary behavior), \textit{Conciseness} (informativeness without unnecessary detail), and \textit{Clarity} (readability and unambiguity).
For the RAG setting, we additionally introduce \textit{Retrieval Utility}, which assesses whether the summary preserves discriminative semantic information for retrieving relevant knowledge.
Each dimension is scored from 1 to 5. 
Consistent with the focus on open-weight evaluation and following Li et al.~\cite{liSFTRLDemystifying2026}, we use GPT-oss-120B as the judge.

\begin{figure}[b!]
    \centering

    \captionsetup[subfigure]{skip=-0.5pt}
    \begin{subfigure}[t]{\columnwidth}
        \centering
        \includegraphics[
            width=\linewidth,
            trim=0 3pt 0 0,
            clip
        ]{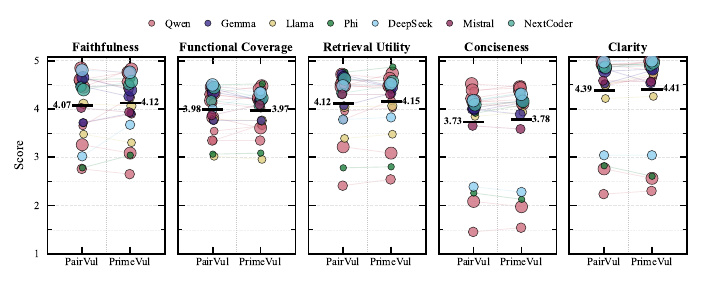}
        \caption{Vul-RAG}
    \end{subfigure}%

    \vspace{-0.25em}

    \begin{subfigure}[t]{\columnwidth}
        \centering
        \includegraphics[
            width=\linewidth,
            trim=0 3pt 0 0,
            clip
        ]{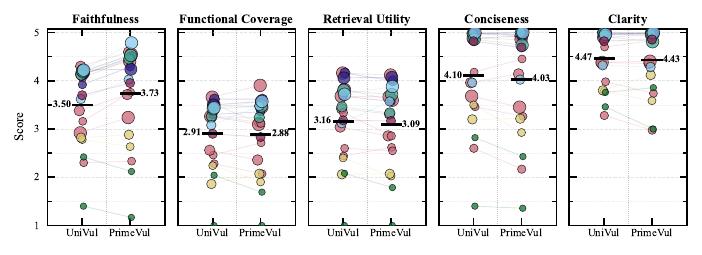}
        \caption{LLM4Vuln\footnote{Due to dependency conflicts in the LLM4Vuln environment, we excluded Qwen3.5-9B and Qwen3.6-27B from its model pool. As a robustness check, we recomputed all means with and without these models and observed no material differences.}}
    \end{subfigure}

    \caption{
        Summary-quality evaluation for system-specific and PrimeVul test probes, with mean scores averaged over the model pool.
        Marker color encodes model family and marker size \sizedots{} increasing parameter count.
    }
    \Description{Two sets of five plots compare LLM-as-a-Judge scores for Faithfulness, Functional Coverage, Retrieval Utility, Conciseness, and Clarity across system-specific and PrimeVul probes for Vul-RAG and LLM4Vuln.}

    \label{fig:summary_quality}





\end{figure}

\newcommand{\rankone}[1]{\cellcolor{diffgreen!30}\textbf{#1}}
\newcommand{\ranktwo}[1]{\cellcolor{diffgreen!15}#1}
\newcommand{\rankthree}[1]{\cellcolor{diffgreen!5}#1}
\begin{table*}[b!]
    \centering
    \fontsize{7pt}{7pt}\selectfont
    \setlength{\tabcolsep}{2.5pt}
    \caption{Inter-judge agreement and score sensitivity for the abstraction-quality evaluation.
    (a)~Ordinal Krippendorff's $\alpha$ and mean pairwise absolute deviation (MAD) across the judges.
    (b)~Mean scores by judge.}
    \label{tab:judge_sensitivity}
    \begin{minipage}[t]{0.35\textwidth}
    \centering
    \textbf{(a) Inter-judge agreement.}\par\vspace{1.5pt}
    \begin{tabular}{@{}lcccc@{}}
        \toprule
        &
        \multicolumn{2}{c}{\textbf{Vul-RAG}} &
        \multicolumn{2}{c}{\textbf{LLM4Vuln}} \\
        \cmidrule(lr){2-3}
        \cmidrule(lr){4-5}
        \textbf{Dimension}
        & \textbf{$\alpha$} & \textbf{MAD}
        & \textbf{$\alpha$} & \textbf{MAD} \\
        \midrule
        Faithfulness        & 0.19 & 0.81 & 0.42 & 0.99 \\
        Functional Coverage & 0.24 & 0.69 & 0.63 & 0.73 \\
        Retrieval Utility   & 0.19 & 0.84 & 0.59 & 0.77 \\
        Conciseness         & 0.16 & 0.79 & 0.28 & 0.82 \\
        Clarity             & 0.38 & 0.48 & 0.36 & 0.59 \\
        \bottomrule
    \end{tabular}

    \vspace{12pt}

    \raggedright
    \noindent System-specific dataset (PairVul for Vul-RAG, UniVul for LLM4Vuln) and PrimeVul probes.
    
    \end{minipage}
    \hfill
    \begin{minipage}[t]{0.63\textwidth}
    \centering
    \textbf{(b) Mean judge scores.}\par\vspace{1.5pt}

    \begin{tabular}{@{}llcccccc@{}}
        \toprule
        &
        &
        \multicolumn{2}{c}{\textbf{GPT-oss-120B}} &
        \multicolumn{2}{c}{\textbf{Qwen2.5-32B}} &
        \multicolumn{2}{c}{\textbf{QwQ-32B}} \\
        \cmidrule(lr){3-4}
        \cmidrule(lr){5-6}
        \cmidrule(lr){7-8}

        & \textbf{Dimension}
        & System & PrimeVul
        & System & PrimeVul
        & System & PrimeVul \\
        \midrule

        \multirow{5}{*}{\rotatebox[origin=c]{90}{\textbf{Vul-RAG}}}
        & \rankthree{Faithfulness}
        & \rankthree{4.07} & \rankthree{4.12}
        & \rankthree{3.92} & \rankthree{3.90}
        & \rankthree{4.58} & \rankthree{4.58} \\

        & \ranktwo{Functional Coverage}
        & 3.98 & 3.97
        & \ranktwo{3.98} & \ranktwo{3.95}
        & \ranktwo{4.63} & \ranktwo{4.63} \\

        & Retrieval Utility
        & \ranktwo{4.12} & \ranktwo{4.15}
        & 3.40 & 3.43
        & 4.39 & 4.28 \\

        & Conciseness
        & 3.73 & 3.78
        & 3.81 & 3.85
        & 4.58 & 4.56 \\

        & \rankone{Clarity}
        & \rankone{4.39} & \rankone{4.41}
        & \rankone{4.40} & \rankone{4.44}
        & \rankone{4.82} & \rankone{4.79} \\

        \midrule

         \multirow{5}{*}{\rotatebox[origin=c]{90}{\textbf{LLM4Vuln}}}
        & \rankthree{Faithfulness}
        & \rankthree{3.50} & \rankthree{3.73}
        & \rankthree{2.92} & \rankthree{2.90}
        & \rankthree{3.71} & \rankthree{3.77} \\

        & Functional Coverage
        & 2.91 & 2.88
        & 2.70 & 2.61
        & 3.45 & 3.35 \\

        & Retrieval Utility
        & 3.16 & 3.09
        & 2.40 & 2.34
        & 3.17 & 3.07 \\

        & \ranktwo{Conciseness}
        & \ranktwo{4.10} & \ranktwo{4.03}
        & \ranktwo{4.28} & \ranktwo{4.37}
        & \ranktwo{4.06} & \ranktwo{3.99} \\

        & \rankone{Clarity}
        & \rankone{4.47} & \rankone{4.43}
        & \rankone{4.28} & \rankone{4.45}
        & \rankone{4.46} & \rankone{4.45} \\

        \bottomrule
    \end{tabular}
    \end{minipage}
\end{table*}

For the evaluation, we retain each system's abstraction prompt and evaluate summaries generated by the different models in the pool for test codes sampled from the original and PrimeVul test sets (cf. Appendix~\ref{app:datasets}). 
Vul-RAG asks the model to extract the purpose and functionality of the code in a specific format, whereas LLM4Vuln requests a short summary (less than three sentences) intended to support search over the vulnerability-knowledge database.

Figure~\ref{fig:summary_quality} visualizes judge-assessed summary quality across the model pool.
Vul-RAG summaries score consistently high in Faithfulness, Functional Coverage, and Retrieval Utility, but lower in Conciseness.
By contrast, LLM4Vuln summaries are generally more concise and clear, but provide less Functional Coverage and Retrieval Utility.
This difference reflects the prompt objectives: Vul-RAG emphasizes purpose and functionality, whereas LLM4Vuln explicitly favors short summaries.
Within each system, mean scores remain largely stable across dimensions and test sets. 
These results suggest that judged abstraction quality is driven more strongly by the system-specific summarization prompt than by the code to be summarized.

Furthermore, several models, including QwQ-32B, Gemma-3-12B, and Qwen2.5-Coder-32B, perform strongly under both systems, indicating that some models perform consistently well across abstraction tasks.
At the same time, several models exhibit pronounced failure cases.
For example, DeepSeek-R1-8B under Vul-RAG produces verbose reasoning traces that primarily reduce Conciseness and Clarity. 
Under LLM4Vuln, Phi-4 represents a more severe failure case, scoring poorly in Faithfulness, Functional Coverage, and Retrieval Utility; manual inspection shows that it returns \texttt{"Code snippet"} for every evaluated query.

\begin{finding}{}
\textbf{Finding \#3}: 
Abstraction quality depends on the system-specific abstraction prompt, with no clear trend attributable to model capabilities; robustness must be verified per pipeline.
\end{finding}

\textit{Judge Sensitivity.}
To assess whether the results depend on the selected LLM judge, we repeat the evaluation with Qwen2.5-32B and QwQ-32B.
Table~\ref{tab:judge_sensitivity} reports ordinal Krippendorff's $\alpha$ and mean pairwise absolute deviation (MAD) across the judges.
Higher $\alpha$ indicates stronger agreement beyond chance, with $\alpha=1$ representing perfect agreement. 
MAD computes the average absolute score difference between judge pairs on the 1-5 scale.
Vul-RAG's $\alpha$ ranges from $0.16$ to $0.38$, which indicates high judge sensitivity.
LLM4Vuln has higher agreement in Functional Coverage ($\alpha=0.63$) and Retrieval Utility ($\alpha=0.59$), but the other dimensions remain judge-sensitive.
Despite differences in absolute scores, the judges preserve the LLM4Vuln abstraction profile, consistently rating Clarity and Conciseness highest, see Tab.~\ref{tab:judge_sensitivity} (b).
For Vul-RAG, judge choice has a stronger effect on individual scores, particularly Retrieval Utility, but the overall ranking of dimensions remains broadly similar.


\subsubsection{Knowledge Retrieval}
\label{sec:rq3-knowledge}
As a high-quality code summary can read well without necessarily helping vulnerability knowledge retrieval, we examine retrieval in practice.
Therefore, we evaluate each knowledge base from two complementary perspectives: \textit{Knowledge Coverage} and \textit{Retrieval Effectiveness}.
Knowledge coverage assesses whether the knowledge base contains knowledge relevant to the vulnerable codes represented in the test set, e.g., CWE type alignment.
Retrieval effectiveness measures whether the retriever can successfully identify this relevant knowledge, quantified using Hit@$k$ and MRR.
Separating these two dimensions allows us to identify retrieval failures caused by missing knowledge alignment and those caused by ineffective retrieval.

\textit{Knowledge Coverage.}
Vul-RAG constructs the knowledge base from historical CVEs, extracting functional semantics, vulnerability causes, and fixing solutions using GPT-3.5~\cite{duVulRAGEnhancingLLMbased2024}.
The system targets $10$ CWEs from the Linux kernel and maintains a dedicated knowledge base and test set for each CWE.
By construction, the knowledge base and test data are aligned.
Vul-RAG also allows for simple regeneration of the knowledge base.
LLM4Vuln uses a \textit{generic} knowledge base generated from CWE descriptions~\cite{sunLLM4VulnUnifiedEvaluation2024}: GPT-4.1 generates $10$ code examples for each of $92$\footnote{The original LLM4Vuln paper reports $86$ CWE categories. Using the publicly available code and configuration, we identified $92$ distinct CWE categories.} CWEs, and further summarizes functionality and root cause.
While this design aims to provide generic vulnerability coverage, the knowledge base is weakly aligned with the curated test data.
Specifically, only $6$ of the $92$ CWEs present in the knowledge base overlap with the $23$ CWEs covered in the UniVul test set, which align with only $60$ of $920$ entries (less than $7\%$). 

\begin{figure*}[b!]
    \centering
    \captionsetup[subfigure]{skip=1pt}

    \begin{subfigure}[t]{\textwidth}
        \centering
        \includegraphics[
            width=\linewidth
        ]{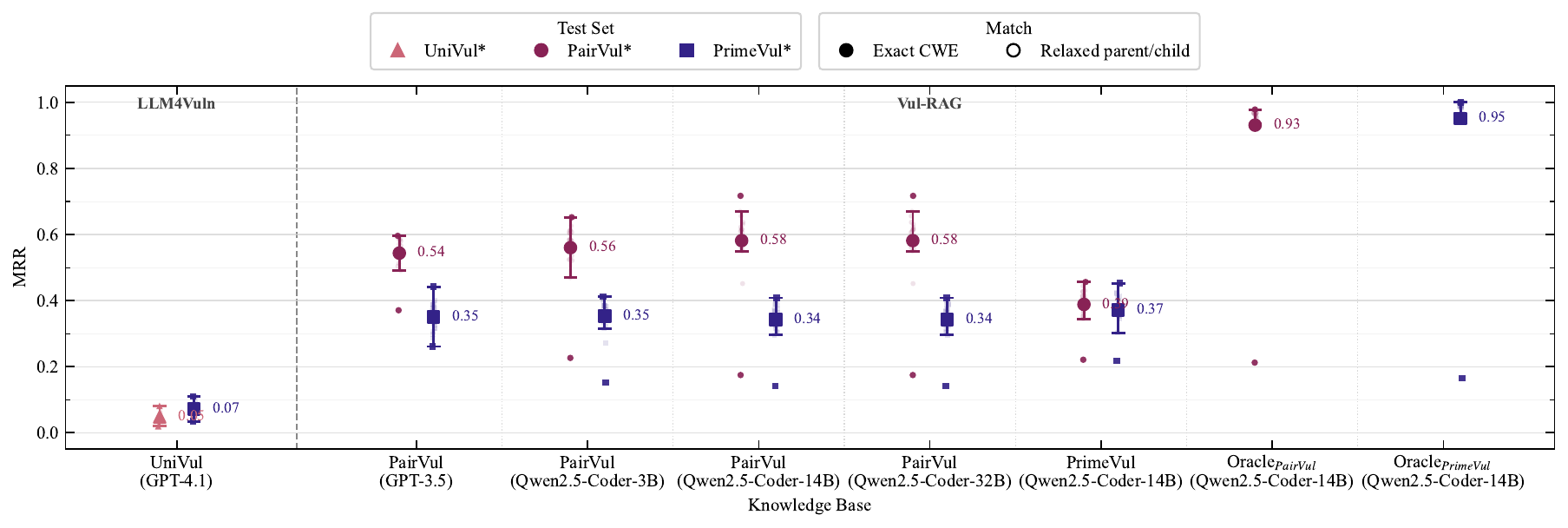}
        \caption{Mean Reciprocal Rank (MRR).}
        \label{fig:retrieval-mrr}
    \end{subfigure}

    \begin{subfigure}[t]{\textwidth}
        \centering
        \includegraphics[
            width=\linewidth
        ]{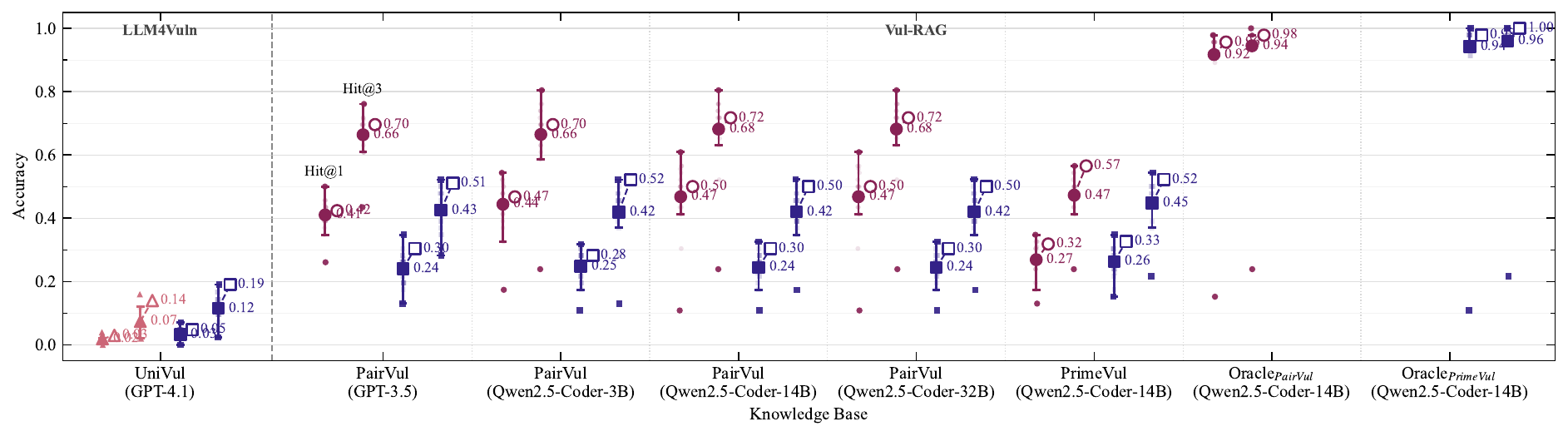}
        \caption{Hit@1 and Hit@3 performance.}
        \label{fig:retrieval-hit}
    \end{subfigure}

    \caption{
        Retrieval effectiveness across knowledge base and test probe combinations for LLM4Vuln and Vul-RAG.
        Knowledge bases are generated from the training sets using the model indicated in parentheses, while marker shape and color identify the test set probe.
        Large filled markers show the mean across input-abstraction models; highlighted points indicate the minimum and maximum observed values; whiskers cover $1.5\times$IQR.
        For Hit@$k$, the line connects the mean under exact ($\bullet$) with the mean under relaxed CWE matching ($\circ$).
    }
    \Description{Retrieval performance for LLM4Vuln and Vul-RAG across different knowledge bases and test probes. The upper panel reports Mean Reciprocal Rank, while the lower panel reports Hit@1 and Hit@3 under exact and relaxed CWE matching.}
    \label{fig:kb-retrieval}

\end{figure*}

\textit{Retrieval Effectiveness.}
For each system, we execute the original retrieval pipeline on summaries generated for the sampled test functions.
We record the ranked knowledge items returned for each query and count a retrieval as relevant when the retrieved item matches the ground-truth CWE of the query function.
To account for the hierarchical structure of CWE, we also consider a relaxed match in which parent or child CWE categories are treated as relevant.
Hit@$k$ therefore measures whether at least one relevant item appears among the top-$k$ results, while MRR captures the rank of the first such match.
Figure~\ref{fig:kb-retrieval} visualizes the results.

For Vul-RAG, retrieval performs moderately well for the PairVul test probe on the original PairVul knowledge base (mean MRR $0.54$, Hit@3 $0.66$), consistent with the alignment analysis above, but substantially worse for the PrimeVul test probe (mean MRR $0.35$, Hit@3 $0.43$).
For LLM4Vuln, retrieval is uniformly weak across both UniVul and PrimeVul test probes (MRR $0.07$, Hit@3 $0.12$ for PrimeVul), in line with its limited alignment between the knowledge base and the test sets.

To establish an upper bound on retrieval, we construct an oracle setting in which the knowledge base is generated from the test set, guaranteeing that a ground-truth item is present in the retrieval pool.
For Vul-RAG, oracle MRR is near-ceiling when the knowledge base and test set are aligned ($0.93$ for PairVul, $0.95$ for PrimeVul), i.e., the retrieval mechanism can recover relevant knowledge effectively when sufficient task-aligned knowledge is available.

\begin{finding}{}
\textbf{Finding \#4}: Across the analyzed pipelines, knowledge base coverage and alignment with the target distribution are crucial for effective retrieval.
Under oracle knowledge, when relevant knowledge is guaranteed to be present, Vul-RAG retrieval reaches near-ceiling effectiveness.
\end{finding}

For Vul-RAG, we further compare knowledge bases regenerated with three models of increasing scale (Qwen2.5-Coder-3B/14B/32B) against the original GPT-3.5-generated PairVul knowledge base.
On the sampled PairVul test set, mean retrieval improves only slightly with increasing generation model scale (MRR: $0.56\rightarrow0.58$; Hit@3: $0.66\rightarrow0.68$).
This trend does not hold on the sampled PrimeVul test set, where retrieval remains flat.
Thus, within the tested models and a fixed knowledge base construction setting, scaling the generation model yields no generalizable additional benefit.

In contrast, retrieval varies substantially more with the model used for input abstraction.
For example, for the original PairVul knowledge base, MRR on the PairVul probe ranges from $0.37$ to $0.60$ across models.
Although the generator can itself affect the quality and alignment of the resulting knowledge base, these results indicate that, among the tested generators, retrieval is more sensitive to how the query is formulated than to which model generated the knowledge base.

\begin{finding}{}
\textbf{Finding \#5}:
Within a fixed knowledge base construction setting, retrieval effectiveness is more sensitive to the query abstraction model than to the knowledge base generation model.
\end{finding}



\subsubsection{Vulnerability Detection}
\label{sec:rq3-detection}

We isolate the detection stage to determine whether the LLM can exploit relevant vulnerability knowledge and translate it into a correct vulnerability prediction.
We conduct two complementary experiments. 
First, we fix Qwen2.5-Coder-14B, which performed best for Vul-RAG in RQ2, as the detection model and vary the Vul-RAG knowledge base to isolate how much of detection performance is determined by the available knowledge.
Second, we fix an oracle knowledge base and vary the detection model to assess how model capabilities bound performance.

\begin{figure*}[t!]
    \centering
    \includegraphics[width=0.95\linewidth,
            trim={0 5 0 5},
            clip]{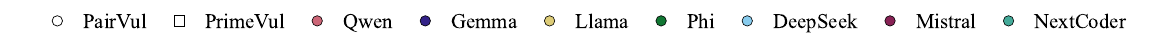}
    \vspace{-0.2em}
    \begin{subfigure}[t]{0.74\textwidth}
        \centering
        \includegraphics[width=\linewidth]{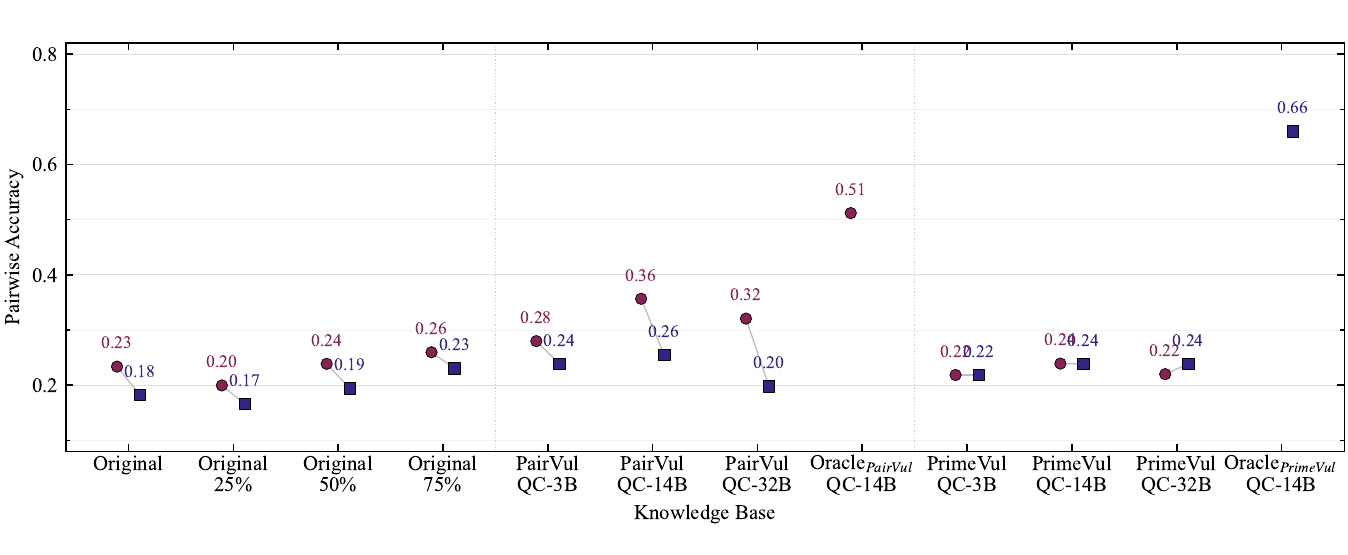}
        \caption{Different knowledge bases.}
        \label{fig:detection_kb}
    \end{subfigure}
    \hfill
    \begin{subfigure}[t]{0.245\textwidth}
        \centering
        \includegraphics[width=\linewidth]{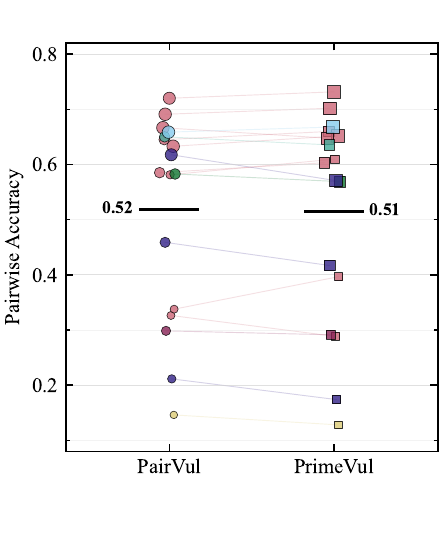}
        \caption{Oracle.}
        \label{fig:detection_oracle}
    \end{subfigure}
    \caption{Vul-RAG detection stage.
    (a) Pairwise Accuracy across knowledge bases with Qwen2.5-Coder-14B fixed as the detector. QC = Qwen2.5-Coder.
    (b) Pairwise Accuracy when oracle knowledge is supplied (for those models that parsed at least 95\% of pairs for both test sets, which excludes $5$ models).}
    \Description{Two panels analyze pairwise accuracy in the Vul-RAG detection stage. The left panel compares different knowledge-base sizes, generators, and source datasets with Qwen2.5-Coder-14B fixed as detector; the right panel compares detector models under oracle knowledge for PairVul and PrimeVul, showing substantial model-dependent variation even with favorable retrieval conditions.}
    \label{fig:detection_analysis}
\end{figure*}

\textit{Sensitivity to the Supplied Knowledge.}
Figure~\ref{fig:detection_kb} visualizes the results.
Sub-sampling the original Vul-RAG knowledge base to 25\%/50\%/75\% of its size yields a consistent upward trend (PairVul:~0.20 $\rightarrow$ 0.24 $\rightarrow$ 0.26; PrimeVul: 0.17 $\rightarrow$ 0.19 $\rightarrow$ 0.23), but accuracy then drops at 100\% size; i.e., the 75\% subset outperforms the full knowledge base.
More knowledge is not uniformly beneficial; we hypothesize this trend reflects either noise from near-duplicate or low-relevance entries diluting top-$k$ retrieval precision, or a higher share of lower-quality entries being surfaced at full scale. 

Regenerating the knowledge base with Qwen2.5-Coder-14B improves pairwise accuracy from $0.23$ to $0.36$, a $52.57\%$ increase.
This gain, however, is not monotonic in generator scale (14B: $0.36$ $>$ 32B: $0.32$ $>$ 3B: $0.28$ on PairVul).
One possible explanation is an implicit alignment between the knowledge generator and the downstream detector: knowledge generated by the same model may better match the preferred representations or abstraction.

Knowledge bases generated from the PrimeVul training set achieve 0.22--0.24 pairwise accuracy regardless of which test set they are evaluated against.
This observation contrasts with the PairVul-derived knowledge bases, which show a clear domain-alignment effect (e.g., pairwise accuracy of $0.36$ on its native test set vs. $0.26$ cross-dataset). 
Together with the higher generation error rate observed for the PrimeVul-based knowledge bases (cf. Appendix~\ref{app:datasets}), the results suggest that the current generation procedure produces less useful knowledge for PrimeVul, possibly due to the longer functions or different vulnerability distribution~\cite{kaniewskiLLM4SVD2026}.

\begin{finding}{}
\textbf{Finding \#6}: In the Vul-RAG pipeline, detection performance is sensitive to the knowledge base, but does not improve monotonically with knowledge base size or generator scale.
The best performance is obtained with Qwen2.5-Coder-14B used for both knowledge generation and vulnerability detection, which points to a generator-detector alignment as an opportunity for improving detection.
\end{finding}

\textit{Detection under Oracle Knowledge.}
To approximate best-case upstream conditions, we use the test set-specific oracle knowledge bases.
Detection is performed on the same test set, with retrieval restricted to knowledge derived from that test set.
This oracle setting intentionally uses test-derived knowledge and is therefore diagnostic rather than deployable; it estimates detector performance when upstream knowledge availability and retrieval are made near-optimal. 

Figure~\ref{fig:detection_oracle} visualizes the results. Mean pairwise accuracy reaches $0.51$ on PrimeVul ($0.66$ for Qwen2.5-Coder-14B), substantially above Vul-RAG under the unified benchmark in RQ2, where Qwen2.5-Coder-14B reaches the best result of $0.22$ pairwise accuracy.
The strongest oracle detectors are larger Qwen-family models, with Qwen3.6-27B reaching $0.73$ on PrimeVul and $0.72$ on PairVul, whereas Llama3.2-3B remains at only $0.13$ pairwise accuracy on PrimeVul and $0.15$ on PairVul despite operating under the same favorable retrieval conditions.
The large spread under oracle knowledge shows that detector capability remains an independent bottleneck.

\begin{finding}{}
\textbf{Finding \#7}: Oracle retrieval does \textit{not} solve vulnerability detection. 
Even under near-optimal oracle conditions, mean pairwise accuracy remains only $0.51$, with the strongest detector reaching $0.73$.
Thus, correctly applying retrieved knowledge remains a substantial bottleneck.
\end{finding}

Across all three analyzed components, the results suggest that improving RAG4SVD requires optimizing the alignment between pipeline stages, rather than simply scaling models or knowledge bases.
Abstraction quality is driven primarily by the system-specific summarization objective (Finding~\#3); retrieval effectiveness depends on query formulation and knowledge base alignment with the target distribution (Findings~\#4 and \#5); and detection performance is influenced more by the alignment of the supplied knowledge than by knowledge base size or generator scale (Finding~\#6).
Even where scaling does help, particularly at the detection stage, it closes only part of the gap: mean pairwise accuracy remains near $0.50$ even under oracle knowledge (Finding~\#7).


\section{Threats to Validity}
\label{sec:threats}

\textit{Internal Validity.}
We re-implemented and adapted existing RAG4SVD systems into a unified benchmark framework. 
Although we preserve system logic, prompts,  and hyperparameters where possible, implementation differences, parser hardening, and model substitutions may affect reproduced results. 
We mitigate these risks by documenting all adaptations in the replication package and repeating experiments where applicable.
All RAG4SVD inference pipelines use locally hosted open-weight models; the only remote dependency is the open-weight GPT-oss-120B judge used for the abstraction-quality analysis in RQ3.

\textit{Construct Validity.}
Pairwise accuracy provides a stricter measure of vulnerable-patched discrimination than instance-level metrics, while excluding configurations with fewer than 95\% parseable predictions may favor systems with more robust output formats. 
We therefore report parsing success in the replication package.
For retrieval, we define relevance by ground-truth CWE agreement, including parent/child relations. 
As CWE agreement does not necessarily imply semantic or vulnerability-specific relevance, we interpret these metrics as measuring knowledge alignment rather than actionable knowledge quality.
The component analysis further relies on sampled test-set probes due to the computational cost of evaluating all model-knowledge base combinations, such that small differences may reflect sampling noise.

\textit{External Validity.}
The findings and conclusions of this work are limited to the systems, datasets, and models considered. 
We study six artifact-complete RAG4SVD systems, use PrimeVul Paired as the common benchmark, and evaluate a documented pool of 3B-32B open-weight models suitable for on-premises deployment.
Results may differ for proprietary models, other programming languages, or vulnerability data distributions. 
Moreover, the component-level analysis focuses on two representative multi-stage pipelines and should therefore be interpreted as evidence for the examined systems rather than as a universal characterization of all RAG4SVD systems.



\section{Conclusion}
\label{sec:conclusion}
In this work, we examined RAG-based software vulnerability detection from the perspectives of reproducibility, cross-system comparability, and component-level attribution.
Across six open-source RAG4SVD systems, we find that reported results do not consistently transfer to open-weight evaluation settings, and that system rankings can change substantially under a unified benchmark.
In the presented component analysis of representative RAG4SVD pipelines, we further dissected input abstraction, knowledge retrieval, and detection to identify where future RAG4SVD systems should focus their engineering effort.
As abstraction quality, e.g., code summarization, is shaped by the system-specific prompting objective, input abstractions should be optimized for retrieval utility rather than conciseness.
As retrieval effectiveness depends on knowledge base coverage and alignment with the target distribution, these properties should be established beforehand, since poor retrieval can limit detection performance.
Finally, oracle retrieval does \textit{not} solve vulnerability detection: oracle knowledge containing ground truth information achieves MRR of $0.93$--$0.95$, yet mean pairwise accuracy remains only $0.51$.

Overall, these findings suggest that progress in RAG4SVD is unlikely to come from scaling individual components, such as models or knowledge bases, alone.
Performance is governed by the alignment across component interfaces, e.g., abstraction to retrieval, knowledge base to target distribution, and retrieved knowledge to detector.
Hence, future evaluations should complement end-to-end detection metrics with component-level evaluations of abstraction quality, knowledge base coverage, retrieval effectiveness, parsing robustness, and downstream knowledge use.


\begin{acks}
This work has been funded by the Federal Ministry of Research, Technology and Space (BMFTR) and the state of Baden-Württemberg (Program: Forschung an HAW, Grant No. 13HAW26PX4), and in part by the German Research Foundation (DFG) under {project-ID 528745080 - FIP 68}. The authors alone are responsible for the content of this paper.
We thank the DACHS data analysis cluster, hosted at Hochschule Esslingen and co-funded by the MWK within the DFG's "Großgeräte der Länder" program, for providing the computational resources necessary for this research.
We also thank Fabian Schmidt for the insightful discussions throughout the experimental process.
\end{acks}



\appendix

\section{Dataset Statistics and Sampling}
\label{app:datasets}
For the component-level analysis in RQ3, we construct sampled test sets that match the CWE distributions of the system-specific reference datasets.
For Vul-RAG, we sample PrimeVul according to the ten CWEs represented in PairVul, selecting five instances per CWE where available.
As PrimeVul contains only one CWE-264 instance, the resulting sample \textit{PrimeVul}*$_{\text{PairVul}}$ contains 46 vulnerable functions.
For LLM4Vuln, we use the full UniVul test set as the reference distribution and sample PrimeVul to match its per-CWE counts where possible.
This yields \textit{PrimeVul}*$_{\text{UniVul}}$ with 42 vulnerable functions across 18 overlapping CWEs.
Tab.~\ref{tab:dataset_stats} summarizes the dataset and probe statistics as well as the knowledge base generation statistics.

\begin{table}[b!]
    \centering
    \fontsize{7pt}{7pt}\selectfont
    \setlength{\tabcolsep}{4pt}
    \caption{Dataset statistics.
    (a)~Train and test datasets, reported with their number of vulnerable functions and CWEs.
    (b)~Test set probes. 
    (c)~Vul-RAG knowledge base generation statistics (PairVul-specific CWE subset).}
    \label{tab:dataset_stats}
    \vspace{-0.75em}
    \begin{minipage}[t]{0.45\textwidth}
    \centering
    \textbf{(a) Dataset statistics.}\par\vspace{1.5pt}
        \begin{tabular}{@{}lrrrr@{}}
            \toprule
            \textbf{Dataset} &
            \multicolumn{2}{c}{\textbf{Train}} &
            \multicolumn{2}{c}{\textbf{Test}} \\
            \cmidrule(lr){2-3}\cmidrule(lr){4-5}
            & \textbf{\#Vuln.} & \textbf{\#CWEs}
            & \textbf{\#Vuln.} & \textbf{\#CWEs} \\
            \midrule
            PrimeVul & 3789 & 111 & 435 & 62 \\
            PairVul  & 2317 & 10  & 586 & 10 \\
            UniVul   & 920  & 92  & 50  & 23 \\
            \bottomrule
        \end{tabular}
        
        \vspace{6pt}
        
    \textbf{(b) Test set probes.}\par\vspace{1.5pt}
        \begin{tabular}{@{}llrr@{}}
            \toprule
            \textbf{System} & \textbf{Probe} & \textbf{\#Vuln.} & \textbf{\#CWEs} \\
            \midrule
            Vul-RAG   & \textit{PairVul}*                         & 46 & 10 \\
            Vul-RAG   & \textit{PrimeVul}*$_{\text{PairVul}}$    & 46 & 10 \\
            LLM4Vuln  & UniVul                                   & 50 & 23 \\
            LLM4Vuln  & \textit{PrimeVul}*$_{\text{UniVul}}$     & 42 & 18 \\
            \bottomrule
        \end{tabular}        
    \end{minipage}
    \hfill
    \begin{minipage}[t]{0.53\textwidth}
    \centering
    \textbf{(c) Knowledge base generation.}\par\vspace{1.5pt}
        \begin{tabular}{@{}llrrr@{}}
            \toprule
            \textbf{Source} &
            \textbf{Generator} &
            \textbf{Expected} &
            \textbf{Generated} &
            \textbf{Errors} \\
            \midrule
            
            \multicolumn{5}{@{}l}{\textit{Training-derived knowledge bases}} \\
            \addlinespace[1pt]
            
            PairVul
            & GPT-3.5 (original)
            & 2317 & -- & -- \\
            
            PairVul
            & Qwen2.5-Coder-3B
            & 2317 & 2312 & 5 \\
            
            PairVul
            & Qwen2.5-Coder-14B
            & 2317 & 2316 & 1 \\
            
            PairVul
            & Qwen2.5-Coder-32B
            & 2317 & 2317 & 0 \\
            
            \addlinespace[1pt]
            
            PrimeVul
            & Qwen2.5-Coder-3B
            & 2398 & 2347 & 51 \\
            
            PrimeVul
            & Qwen2.5-Coder-14B
            & 2398 & 2370 & 28 \\
            
            PrimeVul
            & Qwen2.5-Coder-32B
            & 2398 & 2374 & 24 \\
            
            \midrule
            
            \multicolumn{5}{@{}l}{\textit{Oracle knowledge bases (test-derived)}} \\
            \addlinespace[1pt]
            
            PairVul
            & Qwen2.5-Coder-14B
            & 586 & 585 & 1 \\
            
            PrimeVul
            & Qwen2.5-Coder-14B
            & 247 & 244 & 3 \\
            
            \bottomrule
        \end{tabular}
    \end{minipage}
\end{table}

\section{Model Lineup}
\label{app:model-lineup}
Table~\ref{tab:model-lineup} lists the 22 open-weight models with their checkpoints used in the evaluation. 
The model pool spans multiple model families, parameter scales, and specializations.

\begin{table*}[t]
    \centering
    \fontsize{7pt}{7pt}\selectfont
    \setlength{\tabcolsep}{3pt}
    \caption{Pool of pen-weight models used throughout the benchmark and component analysis. 
    }
    \vspace{-0.75em}
    \label{tab:model-lineup}
        \begin{tabular}{@{}llrllr@{}}
        \toprule
        \textbf{Family} &
        \textbf{Model (Paper Abbreviation)} &
        \textbf{Scale} &
        \textbf{Specialization} &
        \textbf{Full Hugging Face Checkpoint} &
        \textbf{Context Size} \\ \midrule
        
        Qwen
         & \modelmarker{qwen}{0.45mm} Qwen2.5-3B~\cite{qwen25technicalreport}
         & 3.1B & General
         & \texttt{Qwen/Qwen2.5-3B-Instruct}
         & 32K \\
        
         & \modelmarker{qwen}{0.75mm} Qwen2.5-14B~\cite{qwen25technicalreport}
         & 14.7B & General
         & \texttt{Qwen/Qwen2.5-14B-Instruct}
         & 32K \\
        
         & \modelmarker{qwen}{1.05mm} Qwen2.5-32B~\cite{qwen25technicalreport}
         & 32.5B & General
         & \texttt{Qwen/Qwen2.5-32B-Instruct}
         & 32K \\
        
         & \modelmarker{qwen}{0.45mm} Qwen2.5-Coder-3B~\cite{qwen25codertechnicalreport}
         & 3.1B & Code
         & \texttt{Qwen/Qwen2.5-Coder-3B-Instruct}
         & 32K \\
        
         & \modelmarker{qwen}{0.75mm} Qwen2.5-Coder-14.7B~\cite{qwen25codertechnicalreport}
         & 14B & Code
         & \texttt{Qwen/Qwen2.5-Coder-14B-Instruct}
         & 32K \\
        
         & \modelmarker{qwen}{1.05mm} Qwen2.5-Coder-32.5B~\cite{qwen25codertechnicalreport}
         & 32B & Code
         & \texttt{Qwen/Qwen2.5-Coder-32B-Instruct}
         & 32K \\
        
         & \modelmarker{qwen}{0.50mm} Qwen3-4B~\cite{qwen3technicalreport}
         & 4.0B & General
         & \texttt{Qwen/Qwen3-4B}
         & 32K \\
        
         & \modelmarker{qwen}{0.65mm} Qwen3.5-9B~\cite{qwen35blog}
         & 9.0B & General
         & \texttt{Qwen/Qwen3.5-9B}
         & 256K \\
        
         & \modelmarker{qwen}{1mm} Qwen3.6-27B~\cite{qwen36blog}
         & 27.0B & General
         & \texttt{Qwen/Qwen3.6-27B}
         & 256K \\
        
         & \modelmarker{qwen}{1.05mm} QwQ-32B~\cite{qwq32b}
         & 32.5B & General
         & \texttt{Qwen/QwQ-32B}
         & 128K \\
        \addlinespace
        
        Llama
         & \modelmarker{llama}{0.45mm} Llama-3.2-3B~\cite{llama3}
         & 3.2B & General
         & \texttt{meta-llama/Llama-3.2-3B-Instruct}
         & 128K \\
        
         & \modelmarker{llama}{0.60mm} Llama-3.1-8B~\cite{llama3}
         & 8.0B & General
         & \texttt{meta-llama/Llama-3.1-8B-Instruct}
         & 128K \\
        \addlinespace
        
        Phi
         & \modelmarker{phi}{0.5mm} Phi-4-Mini~\cite{phi4minitechnicalreport}
         & 3.8B & General
         & \texttt{microsoft/Phi-4-mini-instruct}
         & 128K \\
        
         & \modelmarker{phi}{0.75mm} Phi-4~\cite{phi4technicalreport}
         & 14.7B & General
         & \texttt{microsoft/phi-4}
         & 16K \\
        \addlinespace
        
        Gemma
         & \modelmarker{gemma}{0.50mm} Gemma-3-4B~\cite{gemma3technicalreport}
         & 4.0B & General
         & \texttt{google/gemma-3-4b-it}
         & 128K \\
        
         & \modelmarker{gemma}{0.70mm} Gemma-3-12B~\cite{gemma3technicalreport}
         & 12.0B & General 
         & \texttt{google/gemma-3-12b-it}
         & 128K \\
        
         & \modelmarker{gemma}{0.95mm} Gemma-3-27B~\cite{gemma3technicalreport}
         & 27.0B & General 
         & \texttt{google/gemma-3-27b-it}
         & 128K \\
        \addlinespace
        
        NextCoder
         & \modelmarker{nextcoder}{0.75mm} NextCoder-14B~\cite{nextcoder}
         & 14.7B & Code 
         & \texttt{microsoft/NextCoder-14B}
         & 32K \\
        
         & \modelmarker{nextcoder}{1.05mm} NextCoder-32B~\cite{nextcoder}
         & 32.5B & Code 
         & \texttt{microsoft/NextCoder-32B}
         & 32K \\
        \addlinespace
        
        DeepSeek
         & \modelmarker{deepseek}{0.60mm} DeepSeek-R1-8B~\cite{deepseekr1}
         & 8.0B & General 
         & \texttt{deepseek-ai/DeepSeek-R1-Distill-Llama-8B}
         & 128K \\
        
         & \modelmarker{deepseek}{1.05mm} DeepSeek-R1-32B~\cite{deepseekr1}
         & 32.5B & General 
         & \texttt{deepseek-ai/DeepSeek-R1-Distill-Qwen-32B}
         & 128K \\
        \addlinespace
        
        Mistral
         & \modelmarker{mistral}{0.58mm} Mistral-7B~\cite{mistral7b}
         & 7.3B & General 
         & \texttt{mistralai/Mistral-7B-Instruct-v0.3}
         & 32K \\
        
        \bottomrule
        \end{tabular}
\end{table*}

\section{Full Reproduction Results}
\label{app:reproducibility}
Tab.~\ref{tab:rep_full} reports the absolute values underlying the RQ1 reproducibility analysis in Sect.~\ref{sec:rq1}. 
For further details, e.g., the calculation of deviations, we refer to the replication package. 

\begin{table*}[t]
\centering
    \caption{Absolute reproduction results for the systems evaluated in RQ1. $\circ$ denotes the reported result, $\bullet_i$ the $i$th reproduction run, and $\overline{\bullet}$ the mean across runs. SVD-Bench provides a seed and yields identical results across all three runs. 
    For Llama-VD, precision, recall, and F1 refer to the vulnerable class. 
    }
    \label{tab:rep_full}
    \fontsize{7pt}{7pt}\selectfont
    \setlength{\tabcolsep}{2.pt}
    \vspace{-0.75em}
\begin{minipage}[t]{0.55\textwidth}
\centering
\textbf{SVD-Bench~\cite{zhangBenchmarkingLargeLanguage2025}}\par\vspace{1.5pt}
    \begin{tabular}{@{}llccccccccc@{}}
        \toprule
        & & \multicolumn{3}{c}{Python}
        & \multicolumn{3}{c}{Java}
        & \multicolumn{3}{c}{JavaScript} \\
        \cmidrule(lr){3-5}\cmidrule(lr){6-8}\cmidrule(lr){9-11}
        Model & 
        & Prec. & Rec. & F1
        & Prec. & Rec. & F1
        & Prec. & Rec. & F1 \\
        \midrule
        
        \multirow{2}{*}{CodeQwen1.5-7B}
        & $\circ$   & 0.11 & 0.52 & 0.18 & 0.11 & 0.48 & 0.18 & 0.19 & 0.56 & 0.29 \\
        & $\bullet$ & 0.08 & 0.20 & 0.11 & 0.08 & 0.25 & 0.12 & 0.20 & 0.50 & 0.28 \\
        \midrule
        
        \multirow{2}{*}{DeepSeek-Coder-7B}
        & $\circ$   & 0.11 & 0.32 & 0.16 & 0.14 & 0.31 & 0.19 & 0.19 & 0.38 & 0.25 \\
        & $\bullet$ & 0.08 & 0.15 & 0.11 & 0.10 & 0.17 & 0.13 & 0.19 & 0.34 & 0.25 \\
        \midrule
        
        \multirow{2}{*}{CodeGemma-7B}
        & $\circ$   & 0.11 & 0.26 & 0.15 & 0.13 & 0.41 & 0.19 & 0.18 & 0.54 & 0.27 \\
        & $\bullet$ & 0.08 & 0.14 & 0.10 & 0.11 & 0.28 & 0.15 & 0.18 & 0.48 & 0.26 \\
        \midrule
        
        \multirow{2}{*}{StarCoder2-7B}
        & $\circ$   & 0.12 & 0.29 & 0.17 & 0.16 & 0.19 & 0.18 & 0.21 & 0.40 & 0.28 \\
        & $\bullet$ & 0.14 & 0.07 & 0.09 & 0.16 & 0.16 & 0.16 & 0.22 & 0.37 & 0.27 \\
        \midrule
        
        \multirow{2}{*}{CodeLlama-7B}
        & $\circ$   & 0.12 & 0.29 & 0.17 & 0.12 & 0.45 & 0.18 & 0.19 & 0.66 & 0.30 \\
        & $\bullet$ & 0.09 & 0.14 & 0.11 & 0.09 & 0.24 & 0.13 & 0.20 & 0.58 & 0.29 \\
        \bottomrule
    \end{tabular}
    
        \vspace{6pt}
        
    \begin{minipage}[t]{0.66\textwidth}
    \centering
    \vspace{3.2pt}\textbf{Llama-VD~\cite{ouchebaraLlamabasedSourceCode2025}} - Llama-3.1-8B\par\vspace{1.5pt}
        \begin{tabular}{@{}lcccccccc@{}}
            \toprule
            & \multicolumn{4}{c}{Big-Vul}
            & \multicolumn{4}{c}{PrimeVul} \\
            \cmidrule(lr){2-5}\cmidrule(lr){6-9}
            & Acc. & Prec. & Rec. & F1
            & Acc. & Prec. & Rec. & F1 \\
            \midrule
            $\bullet_1$
            & 0.69 & 0.70 & 0.65 & 0.68
            & 0.66 & 0.68 & 0.63 & 0.65 \\
            $\bullet_2$
            & 0.70 & 0.71 & 0.67 & 0.69
            & 0.67 & 0.68 & 0.64 & 0.66 \\
            $\bullet_3$
            & 0.70 & 0.71 & 0.67 & 0.69
            & 0.67 & 0.68 & 0.63 & 0.66 \\
            \midrule
            $\circ$
            & 0.70 & 0.71 & 0.67 & 0.69
            & 0.67 & 0.68 & 0.64 & 0.66 \\
            $\overline{\bullet}$
            & 0.70 & 0.71 & 0.66 & 0.69
            & 0.67 & 0.68 & 0.63 & 0.66 \\
            \bottomrule
        \end{tabular}
    \end{minipage}
    \hfill
    \begin{minipage}[t]{0.33\textwidth}
    \centering
    \textbf{Vul-RAG~\cite{duVulRAGEnhancingLLMbased2024}} - Qwen2.5-Coder-32B\par\vspace{1.5pt}
        \begin{tabular}{@{}lccc@{}}
            \toprule
            & \multicolumn{3}{c}{PairVul} \\
            \cmidrule(lr){2-4}
            & Pair. Acc. & Rec. & Prec. \\
            \midrule
            $\bullet_1$    & 0.25 & 0.58 & 0.58 \\
            $\bullet_2$    & 0.25 & 0.59 & 0.60 \\
            $\bullet_3$    & 0.27 & 0.60 & 0.60 \\
            \midrule
            $\circ$ & 0.26 & 0.59 & 0.61 \\
            $\overline{\bullet}$
            & 0.26
            & 0.59
            & 0.59 \\
            \bottomrule
        \end{tabular}
    \end{minipage}
\end{minipage}
\hfill
\begin{minipage}[t]{0.4\textwidth}
\centering
\textbf{LLM4Vuln~\cite{sunLLM4VulnUnifiedEvaluation2024}}\par\vspace{1.5pt}
    \begin{tabular}{@{}llcccccc@{}}
        \toprule
        & & \multicolumn{3}{c}{C/C++}
        & \multicolumn{3}{c}{Java} \\
        \cmidrule(lr){3-5}\cmidrule(lr){6-8}
        Model &
        & Prec. & Rec. & F1
        & Prec. & Rec. & F1 \\
        \midrule
        
        \multirow{5}{*}{QwQ-32B}
        & $\bullet_1$    & 0.21 & 0.19 & 0.14 & 0.08 & 0.02 & 0.11 \\
        & $\bullet_2$    & 0.18 & 0.16 & 0.13 & 0.03 & 0.01 & 0.04 \\
        & $\bullet_3$    & 0.19 & 0.17 & 0.13 & 0.04 & 0.01 & 0.06 \\ \cmidrule{2-8}
        & $\circ$ & 0.26 & 0.33 & 0.29 & 0.14 & 0.12 & 0.13 \\
        & $\overline{\bullet}$ & 0.19 & 0.17 & 0.13 & 0.05 & 0.01 & 0.07 \\
        \midrule
        
        \multirow{5}{*}{Llama-3.1-8B}
        & $\bullet_1$    & 0.12 & 0.12 & 0.11 & 0.04 & 0.03 & 0.03 \\
        & $\bullet_2$    & 0.19 & 0.19 & 0.17 & 0.07 & 0.06 & 0.05 \\
        & $\bullet_3$    & 0.15 & 0.15 & 0.14 & 0.05 & 0.04 & 0.04 \\ \cmidrule{2-8}
        & $\circ$ & 0.09 & 0.17 & 0.12 & 0.05 & 0.09 & 0.06 \\
        & $\overline{\bullet}$ & 0.15 & 0.15 & 0.14 & 0.05 & 0.04 & 0.04 \\
        \midrule
        
        \multirow{5}{*}{Phi-3-Mini-128K}
        & $\bullet_1$    & 0.21 & 0.17 & 0.21 & 0.08 & 0.08 & 0.08 \\
        & $\bullet_2$    & 0.17 & 0.15 & 0.18 & 0.08 & 0.08 & 0.08 \\
        & $\bullet_3$    & 0.19 & 0.16 & 0.19 & 0.08 & 0.08 & 0.08 \\ \cmidrule{2-8}
        & $\circ$ & 0.10 & 0.20 & 0.13 & 0.07 & 0.14 & 0.10 \\
        & $\overline{\bullet}$ & 0.19 & 0.16 & 0.19 & 0.08 & 0.08 & 0.08 \\
        \bottomrule
    \end{tabular} 
\end{minipage}
\end{table*}

\bibliographystyle{ACM-Reference-Format}
\bibliography{literature}


\end{document}